\documentclass[journal=jacsat,manuscript=article]{achemso}
\usepackage[utf8]{inputenc} 

\usepackage[font=small, labelfont=bf, justification=justified,singlelinecheck=false]{caption}
\usepackage{enumerate}
\usepackage{amsmath}
\usepackage{amssymb}

 \usepackage{caption}
 \usepackage{subcaption}
\usepackage{ifpdf}
\usepackage{rotating}
\usepackage{alltt}
\usepackage{color}
\usepackage[usenames,dvipsnames]{xcolor}

\definecolor{lgray}{gray}{0.7}
\definecolor{llgray}{gray}{0.5}
\definecolor{lllgray}{gray}{0.3}

\usepackage{setspace}

\newcommand*{\xlineshort}[1][1.2em]{\rule[0.4ex]{3.5pt}{0.5pt}}

\newcommand*{\xdash}[1][1.2em]{\rule[0.4ex]{2.5pt}{0.5pt} \rule[0.4ex]{2.5pt}{0.5pt}}
\newcommand*{\xdashthick}[1][1.2em]{\rule[0.4ex]{3.5pt}{1.5pt} \rule[0.4ex]{3.5pt}{1.5pt}}
\newcommand*{\xdashvthick}[1][1.2em]{\rule[0.4ex]{4.5pt}{2.5pt} \rule[0.4ex]{4.5pt}{2.5pt}}

\newcommand{\eq}[1]{ Eq.\ (\ref{#1})}

\DeclareMathOperator*{\Pressuretype}{P}

\DeclareMathOperator*{\StressVIRIAL}{{\Pressuretype}}

\DeclareMathOperator*{\StressIKONE}{\boldsymbol{\Pressuretype\limits^{\scriptscriptstyle{IK1}}}}

\DeclareMathOperator*{\PressureVA}{{\Pressuretype\limits^{\scriptscriptstyle{V\!A}}}}

\newcommand{\ie}{${\it i.e.\ }$}

\DeclareMathOperator*{\intV}{\Delta V}

\DeclareMathOperator*{\MDvel}{\boldsymbol{v}_{\mathnormal{i}}}
\DeclareMathOperator*{\MDpvel}{{\boldsymbol{p}}_{\mathnormal{i}}}

\DeclareMathOperator*{\CFDvel}{\boldsymbol{u}}

\DeclareMathOperator*{\Fij}{\boldsymbol{F}_{ij}}
\DeclareMathOperator*{\Fijrij}{\boldsymbol{r}_{ij} \Fij}

\title{A Spatio-Temporal Generalisation of Green Kubo}

\author{E. R. Smith}
\email{edward.smith05@imperial.ac.uk}
\affiliation{Department of Mechanical and Aerospace Engineering, Brunel University London, Uxbridge, United Kingdom} 
\author{D. Dini}
\email{d.dini@imperial.ac.uk}
\affiliation{%
Department of Mechanical Engineering,\\
Imperial College London, Exhibition Road, 
South Kensington, London SW7 2AZ.
United Kingdom,\\
}%
\author{D.M. Heyes}
\email{d.heyes@imperial.ac.uk}
\affiliation{%
Department of Mechanical Engineering,\\
Imperial College London, Exhibition Road, 
South Kensington, London SW7 2AZ.
United Kingdom,\\
}%

\begin{document}

\date{\today}
\begin{abstract}

The Green-Kubo (GK) method which is used to obtain the shear viscosity of a model liquid in equilibrium molecular dynamics (MD) often requires long simulation times to obtain acceptable statistics.
This work extends the GK approach to include spatial correlations.
The total shear stress in the GK expression is split into a grid of its components in contiguous volumes.
In doing this, the autocorrelation of the total system stress can be rewritten as the cross-correlation of the subvolume stresses. 
This provides a novel real-space insight into the liquid structure, something previously only treated in Fourier space by liquid-state theory.
A novel travelling wave like structure is exposed that is shown to be well fitted by a leading Gaussian pulse added to a second negative Gaussian for the bounce back.
The fitting parameters include the wave position, magnitude and width which show deep insights into the liquid property.
The wave moves at the speed of sound over short times before decreasing in speed with the square root of time, while the wave packet spreads out also as the square root of time.
The magnitude is decreasing with form $t^{-3/2}$, a result with strong significance in the history of MD simulation.
These fitted forms mean the entire spatial temporal response of the liquid can be modelled in closed form by fitting to data, with short term requiring MD and the long time and distances approximated by the Gaussian form.
By limiting the correlation to localized interactions the longer range contribution,
which essentially only contribute to the noise, can be eliminated.
This looks like a promising approach to model viscosity by taking short MD runs and fitting the long time behaviour.

\end{abstract}

\maketitle

\section{Introduction}
\label{sec:Intro}

Obtaining the viscosity of a liquid has long been an important application of molecular dynamics (MD) simulation.
As the most fundamental classical methodology, MD provides a complete picture of 
the fluid structure.
The viscosity is an output of the simulation, observed from the average behaviour of the molecules as they evolve over time.
Once obtained, viscosity is often the only empirical coefficient needed to describe the evolution of Newtonian fluids in computational fluid dynamics (CFD) simulation.
More recently, as computers and software advance, simulations increasingly 
aim to model non-Newtonian fluids by inputting the viscosity directly into a CFD solver as part of a coupled model \citep{HMM}, either by running prior parameterising simulations or generated `on the fly' in parallel.
Such models 
offer potential advantages in modelling
a range of 
practical systems where the accuracy of outcome is limited 
by the quality of the inputted rheological constitutive equation.
As a result, the efficient calculation of viscosity becomes an 
increasingly more important issue.\\

The first MD simulations of the (Newtonian) shear viscosity by MD were for hard spheres,~\cite{alderwainwright1970} 
and the Lennard-Jones,~\cite{LevesqueVerlet1970}
liquid, carried out at more or less the same time using the Einstein-Helfand (EH) and formally equivalent 
Green-Kubo (GK) routes,~\cite{Hess,Evans2001Hess,Evans2003Hess} respectively. The EH 
approach involves essentially analysing the fluctuations
of shear stress in block averages of different length, whereas the GK method involves calculating 
and integrating with time the shear 
stress autocorrelation function (which has an intrinsic interest 
in its own right). Both  routes are based on the shear 
stress fluctuations and essentially involve the computation 
of the statistical inefficiency factor, $s$, (or `correlation time', $\tau_{c}$) usually discussed in the context of 
evaluating accurate standard error values in simulation property 
averages.~\cite{allentildesleybook} Since then
both EH and GK have been applied to a wide range of molecular systems. The statistical 
accuracy and the factors that improve it, of the GK method have been the subject or many publications.
Most of these have been concerned with the convergence of the GK correlation function and its
integral with time.~\cite{ZhangOtani2015,NevinsSpera2007} The relative error in the transport coefficient scales as
$\simeq \sqrt{\tau_{c}/t_{sim}}$ where $t_{sim}$ is the 
duration of the simulation.~\cite{JonesMandadapu2012,KimHanKim2018}. A resolution of the 
shear stress autocorrelation function, $C_{s}(t)$, or SACF
into molecule-centric components, either as the self- and cross-terms, 
or alternatively $2-$, $3-$ and $4-$body 
contributions,~\cite{StassenSteele1994} has shown that these 
individual components decay more slowly with time than the total correlation function
because of a cancellation in the summation. \\


In comparison, much less attention has been given to the length-scale issues
in regard to the statistical efficiency of the viscosity evaluation.
It has been found that the system size-dependence 
of the shear viscosity is rather weak
even near the triple point, certainly compared to the self-diffusion coefficient.~\cite{heyesCass2007}
Although Petravic identified,~\cite{petravic2004} that for small systems the liquid 
can sustain on a simulation time scale a
shear stress, which was related to the cooperatively rearranging region of the Adams-Gibbs
theory of glass formation.
The origin of the viscosity in supercooled model liquids has been explored by MD using the atomic 
stress component
of the stress tensor, in the form of its self and cross component correlation functions.
These studies have shown there is a connection between the viscosity and the shear stress waves
which are both slowly decaying in space and time.~\cite{Chowdhury2016,LevashovMorris2013,LevashovMorris2017}

However, it was found that the calculation of viscosity was inherently non-local, often requiring many layers of intermolecular interactions.
Non-equilibrium MD (NEMD) is another route to the shear viscosity, in which a shear velocity 
profile is imposed on the fluid. Both wall-driven,~\cite{AshurstHoover},
sinusoidal forms in a periodic system,~\cite{goslingsinger} and synthetic  algorithm (e.g., SLLOD,~\cite{evansmorriss1984,Ladd}) combined
with Lees-Edwards boundary conditions approaches have been used.~\cite{LeesEdwards}
NEMD lends itself to the computation of 
local quantities
and a range of mathematical prescriptions have been developed to calculate the pressure tensor with spatial resolution.
These various methods all stem from the seminal work of Irving and Kirkwood,~\citet{Irving_Kirkwood}, 
which involves using a kernel function that is integrated over a volume in space or across a plane.
Arguably the only meaningful definition of shear stress
is one which measures force aligned with a reference plane.
The plane must be chosen so that it is independent of the positions of the molecules themselves.
In recent work, we showed that the presence of a particle 
at the center of a volume will change of the measured stress,~\cite{spherepressure} an observation which suggests that decomposing the stress by spatial volume rather than with
a molecule at the center of the volume is preferable. This is one of the main conclusions
to be derived from this work.
By incorporating the local definition of shear stress within the equilibrium Green-Kubo formulations, we propose a potentially more efficient method of measuring the shear viscosity of molecular systems, by excluding contributions from uncorrelated subvolumes.

This work is organised as follows, in section \ref{sec:Theory} the basic equations are stated. In section \ref{sec:Simulation}, the details of the molecular model are set out. In section \ref{sec:Results}, the results and discussion are presented followed by a process of fitting the behaviour to a reduced model, before concluding remarks in section \ref{sec:Conclusions}.

\section{Theory}
\label{sec:Theory}

In this section, a method of getting a spatial decomposition of viscosity is presented by dividing the Green Kubo approach into subvolmes inside the domain \citep{Green, Kubo}
The Green-Kubo \citep{Green, Kubo} relations provide a way to determine the viscosity in an equilibrium system from the autocorrelation of shear stress as follows,
\begin{align}
\mu = \frac{V}{k_B T} \int_0^\infty \large\langle  P_{xy} (\tau)  P_{xy}(0) \large\rangle d \tau = \frac{V}{k_B T} \int_0^\infty \mathcal{C}(\tau) d \tau
\label{GreenKubo}
\end{align}
where $\StressVIRIAL{}_{xy} $ is the shear stress, $V$ the volume of the system, $T$ the temperature and $k_B$ Boltzmann's constant. 
The angular brackets denote an average over an ensemble of systems but the ergodic hypothesis means we can typically use a range of initial starting times for the autocorrelation integral instead. The notation $\mathcal{C}$ is introduced for the integrand here.

The pressure tensor in the entire domain is given by the Virial \citep{Clausius} expression generalised to a tensor \citep{Parker},
\begin{align}
\boldsymbol{\StressVIRIAL} = \frac{1}{V} \left[\sum_{i=1}^N \frac{\MDpvel \MDpvel}{m_i} + \sum_{i,j}^N \Fijrij \right]
\label{virial}
\end{align}
where $\MDpvel/m_i  = \MDvel - \CFDvel$ is peculiar velocity and $i,j$ indicates a double sum over all $i$ and $j$ with $i \ne j$.
This expression is only valid for an entire domain, but following the work of \citet{Irving_Kirkwood_1950}, the pressure at a local point in space can obtained.
This form of local pressure is non-unique, dependent on the volume chosen and the interaction path between any interacting molecules.
In this work a uniform grid of cuboids is used, so measured stress depends on the chosen interaction path.
The simplest path is known as the IK1, a truncation of the IK expansion of Dirac delta function to just the terms per molecules. This essentially decomposes the pressure in half assigning it to the location of each molecules, so local pressure is simply.
\begin{align}
\boldsymbol{\StressIKONE}_I = \frac{1}{\intV} \sum_{i=1}^N \left[  \frac{\MDpvel \MDpvel}{m_i} + \sum_{j \ne i}^N \Fijrij  \right] \vartheta_i^I
\label{virial}
\end{align}
where $\vartheta_i^I$ is a function which is one if molecule $i$ is inside volume $I$ and zero otherwise with $\intV$ the cell volume.
Adding up these stresses per molecule for all cells in the domain returns the Virial expression.
 
A more formal treatment of the \citet{Irving_Kirkwood_1950} expansion gives an intergral between molecules. Assuming this is a simple linear interaction path yields the volume average (VA) pressure, which for any individual volume is given by,
\begin{align}
\boldsymbol{\PressureVA}_I = \frac{1}{\intV} \left[\sum_{i=1}^N \frac{\MDpvel \MDpvel}{m_i} \vartheta_i^I + \sum_{i,j}^N \Fijrij \ell_{ij}^I \right]
\label{VA}
\end{align}
where $\ell_{ij}^I$ is a function which obtaines the fraction of the interaction line inside volume $I$.
The relation between Virial and volume average is also exact, the sum of VA for all cells is the Virial. It is helpful to think of the VA as simply a book-keeping technique which divides the total Virial pressure into local contributions by partitioning the pressure contributions based on the location of the particles and their line of interaction.
There is a notable differences with the VA form of pressure compared to the IK1, this captures the fraction of the inter-molecular interactions which cross cells between molecules.
As a result, the pressure distribution is less dependent on the actual location of molecules themselves.
In a homogenous fliuid, the IK1 and VA give identical results but they differ in heterogenous systems. 
A classic example of this is near walls in non-equilibrium systems where the molecules tend to sit in energetically favourable locaitons due to interaction with the solid lattice, seen as a density stacking on binned plots and a continuation of the solid-like structure into the liquid in visualisations of the molecules.
In these cases, the pressure distribution from the IK1 is not flat in an equilibrium system, an artefact which violates momentum balance $\boldsymbol{\nabla} \cdot \boldsymbol{P} \ne 0$.
As the domain is filled with a contiguous grid of cuboidal volumes, the sum of all volumes returns the Virial exactly in both IK1 and VA cases,
\begin{align}
\boldsymbol{\StressVIRIAL} V  = \sum_{I=1}^{N_{x_{cell}}} \sum_{J=1}^{N_{y_{cell}}} \sum_{K=1}^{N_{z_{cell}}} \boldsymbol{\StressIKONE{}}^{IJK} \!\!\!\! \intV = \sum_{I=1}^{N_{x_{cell}}} \sum_{J=1}^{N_{y_{cell}}} \sum_{K=1}^{N_{z_{cell}}} \boldsymbol{\PressureVA{}}^{IJK} \!\!\!\! \intV 
\label{virial_to_VA}
\end{align}
where the domain is divided into tessellating cubic volumes with $N_{x_{cell}}$ in $x$, $N_{y_{cell}}$ in $y$ and $N_{z_{cell}}$ in $z$. 
For notational simplicity, we reduce the 3 spatial indices to a single one which can be though of as successive shells in three dimensional space,
The shorthand for sum over all cells is used with $N_{cell} = N_{x_{cell}} N_{y_{cell}} N_{z_{cell}}$.

Combining the definition of the autocorrelation in \eq{GreenKubo} of virial pressure with the sum of stress in all volumes \eq{virial_to_VA}, results in,
\begin{align}
\mu & = \frac{V}{ k_B T} \int_0^\infty \left\langle   \frac{\Delta V }{V} \sum_{I=1}^{N_{{cell}}} \PressureVA{\!\!}_{{xy}}^{\; I} (\tau)  \frac{\Delta V }{V} \sum_{L=1}^{N_{{cell}}} \PressureVA{\!\!}_{{xy}}^{\; L} (0) \right\rangle d \tau \nonumber \\
& =  \frac{(\Delta V)^2}{ k_B V T} \int_0^\infty \left\langle  \sum_{I=1}^{N_{{cell}}} \sum_{L=1}^{N_{{cell}}}  \PressureVA{\!\!}_{{xy}}^{\; I} (\tau) \PressureVA{\!\!}_{{xy}}^{\; L} (0) \right\rangle d \tau
\label{mu_spatial_sum}
\end{align}
This product of two sums over all cells gives $N_{cell}^2$ terms. 
We can sort the various contributions into a self correlation (the autocorrelation in a cell) and a series of cross correlations between adjacent cells.

\begin{figure}
\includegraphics[width=0.9\textwidth]{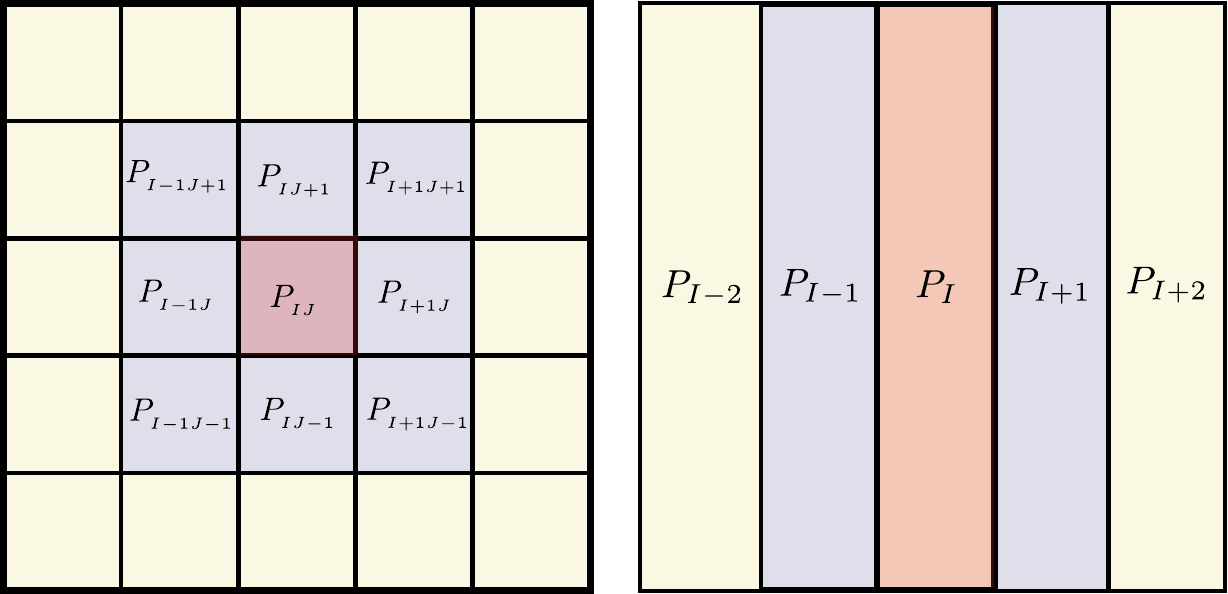}
\caption{Schematic showing the cells in layers which are used in the calculation of the autocorrelation with $a)$ the 3D grid of cells in a periodic box and $b)$ a 1D version with cells just in the x direction with a periodic box in all directions.}  
\label{schematic_layered_averages}
\end{figure}

To illustrate this, we consider a 2D case focused on the correlation of all other cells with a central cell with index $I, J$.
This is shown graphically in Fig. \ref{schematic_layered_averages}, with only the first layer annotated.
As we are only interested in shear pressure for viscosity calculations, we use the shorthand  $\PressureVA{\!\!}_{_{IJ}}= \PressureVA{}_{xy}^{IJ} $, so that,
\begin{align}
 \sum_{L=1}^{N_{x_{cell}}} \sum_{M=1}^{N_{y_{cell}}} \PressureVA{\!\!}_{_{LM}} (\tau)   \PressureVA{\!\!}_{_{IJ}} (0)  = \underbrace{  \PressureVA{\!\!}_{_{IJ}} (\tau) \PressureVA{\!\!}_{_{IJ}} (0) }_{\textrm{Autocorrelation}} + \underbrace{ \vphantom{  \PressureVA{\!\!}_{_{IJ}} }  \mathcal{C}_{_1} + \mathcal{C}_{_2} + \mathcal{C}_{_3} + ...}_{\textrm{Cross Correlation Terms}} 
\end{align}
where the successive shells of volumes around cell $I,J$ is denoted with $\mathcal{C}_{_0}$ for the autocorrelation, $\mathcal{C}_{_1}$ for layer 1, the correlation with the surrounding $8$ volumes in 2D ($26$ in 3D), and $\mathcal{C}_{_2}$ for layer 2, the next $16$ in 2D ($99$ in 3D) and so on.
The cross correlation terms for $\mathcal{C}_{_1}$ for the 2D case would therefore be as follows,
\begin{align}
\mathcal{C}_{_1} =  
    \PressureVA{\!\!}_{_{I-1   \; J-1}} (\tau)  \PressureVA{\!\!}_{_{IJ}} (0) 
\; + \;  \PressureVA{\!\!}_{_{I-1   \; \;J\; }} (\tau) \PressureVA{\!\!}_{_{IJ}} (0) 
\; + \;   \PressureVA{\!\!}_{_{I-1   \; J+1}} (\tau) \PressureVA{\!\!}_{_{IJ}} (0) 
 \nonumber \\  
\; + \;   \PressureVA{\!\!}_{_{\;I\; \; J-1}} (\tau) \PressureVA{\!\!}_{_{IJ}} (0) 
\;\;\;\;\;\;\;\;\;\;\;\;\;\;\;\;+ \;\;\;\;\;\;\;\;\;\;\;\;\;\;\;\;\;\;\;
   \PressureVA{\!\!}_{_{\;I\; \; J+1}} (\tau) \PressureVA{\!\!}_{_{IJ}} (0) 
\nonumber \\  
\; + \;   \PressureVA{\!\!}_{_{I+1 \; J-1}} (\tau) \PressureVA{\!\!}_{_{IJ}} (0)  
\; + \;   \PressureVA{\!\!}_{_{I+1 \; \;J\;}} (\tau)  \PressureVA{\!\!}_{_{IJ}} (0)
\; + \;   \PressureVA{\!\!}_{_{I+1 \; J+1}} (\tau)  \PressureVA{\!\!}_{_{IJ}} (0)
\end{align}
which is shown schematically in Fig \ref{schematic_layered_averages} $a)$.
Note that periodic boundaries are applied in all directions and so are used in taking cross correlations that span boundaries.
In practice, due to the $N_{cell}^3$ scaling of three dimensional cross-correlations, grouping layers together or, as is done in this work, correlating in only one dimension is often essential and this 1D slab based system is shown schematically in Figure \ref{schematic_layered_averages} $b)$. 
In the one dimensional case, this expression is much simpler and can be though of as a successive sum over increasing layers,
\begin{align}
 \sum_{L=1}^{N_{x_{cell}}} \PressureVA{\!\!}_{_{L}} (\tau)   \PressureVA{\!\!}_{_{I}} (0)  = \PressureVA{\!\!}_{_{I}} (0) \Bigg( \underbrace{  \PressureVA{\!\!}_{_{I}} (\tau)  }_{\textrm{Layer 0}} + \underbrace{ \vphantom{  \PressureVA{\!\!}_{_{IJ}} }  
 \PressureVA{\!\!}_{_{I-1}} (\tau)  + \PressureVA{\!\!}_{_{I+1}} (\tau) }_{\textrm{Layer 1}} 
+ \underbrace{ \vphantom{  \PressureVA{\!\!}_{_{IJ}} }  
 \PressureVA{\!\!}_{_{I-2}} (\tau) + \PressureVA{\!\!}_{_{I+2}} (\tau) }_{\textrm{Layer 2}}
+ \dots 
\end{align}
aslo shown schematically in Fig \ref{schematic_layered_averages} $b)$.
Note these correlation terms per layer are taken over an ensemble average, using many time origins averaged to give each term.

In order to make sense of this form, we use a intuitive argument here to show this can be interpreted as a spatial correlation equivalent to the time integral used in the original Green-Kubo correlation.
As the choice of cellsize is arbitrary, we can make them as small as possible to get smoother results for the spatial dependence of spatial correlation of pressure.
Taking the limit that the cellsize $\Delta V \to 0$, the sum can be used as the definition of a Riemann integral, so in the limit of zero volume the sum over all cells surrounding cell $M$ tends to a spatial integral,
\begin{align}
 \lim_{\Delta V \to 0} \sum_{J=1}^{N_{cell}}  \PressureVA{\!\!}_{_{J}} (\tau)   \PressureVA{\!\!}_{_{M}} (0) \Delta V    = \int_0^{\infty}  \PressureVA{}_{xy} (r, \tau)  \PressureVA{}_{xy}(0, 0)  dr
\label{Riemann}
\end{align}
where in the limit of zero volume, the control volume function tends to a Dirac delta form $\vartheta_i \to \delta(r - r_i)$ and the VA pressure tensor tends to the Irving Kirkwood form, valid at a point \citep{Smith_et_al12}.
The angular brackets denote averaging over an ensemble of systems, or in practice the correlation over $N_e$ distinct time origins \citep{Allen_Tildesley} $\langle A(t) A(0) \rangle_t \approx 1/N_e \sum_e^{N_e} A(t+e) A(e)$ where we add the subscript $t$ to emphasise this is time based averaging.
We can extend the concept of multiple origin averaging to include the multiple spatial cell origins as the starting point for the spatial integral, \ie 
\begin{align}
\langle \PressureVA{}_{xy} (r, \tau)  \PressureVA{}_{xy}(0, 0) \rangle_r \approx  \frac{1}{N_{cell} } \sum_{M=1}^{N_{cell}} \PressureVA{}_{xy} (r+\Delta x M, \tau)  \PressureVA{}_{xy}(\Delta x M, 0) 
\label{space_ensemble}
\end{align}
Where the subscript $r$ is used to denote a spatial average on the angular brackets.
If we note that $V/\Delta V=N_{cells} $, then using the definition of the integral from Eq \ref{Riemann} and ensemble definition of Eq \ref{space_ensemble}, the Green Kubo expression Eq \ref{mu_spatial_sum} can be rewritten as,
\begin{align}
\mu & =   \lim_{\Delta V \to 0} \frac{1}{k_B T} \int_0^\infty \left\langle  \frac{1}{ N_{cells}}  \sum_{I=1}^{N_{{cell}}} \sum_{L=1}^{N_{{cell}}}  \PressureVA{\!\!}_{{xy}}^{\; I} (\tau) \PressureVA{\!\!}_{{xy}}^{\; L} (0) \Delta V \right\rangle_t d \tau  \nonumber \\
& = \frac{1}{k_B T} \int_0^\infty  \int_0^{\infty} \left\langle \PressureVA{}_{xy} (r, \tau)  \PressureVA{}_{xy}(0, 0)  \right\rangle_{r,t}  dr d\tau= \frac{1}{k_B T} \int_0^\infty  \int_0^\infty \mathcal{C}(r,t) dr d \tau.
\end{align}
Which is the final form of the Spatial Green Kubo (SGK) correlation.
This is therefore a two dimensional correlation, in both space and time, which we denote with the integrand function of both $\mathcal{C}(r,t)$.
This highlights one immediate advantage of this spatial approach, the statistics can be improved by averaging over both a shifting-spatial origins and varying time-origin.
It is this spatial dependence we utilise in the next section, firstly to gain insights into exactly how viscosity emerges from the local molecular structure.
However, more importantly for improving statistics, we can also truncate both spatial and temporal correlations to minimise uncertainty in our SGK measurement.

\section{Simulation}
\label{sec:Simulation}

This study uses a periodic box of molecular run using MD code Flowmol, which has been validated for MD simulation \citep{Smith_Thesis} and VA stress measurements \citep{CV_paper} in previous work.
The simulations use the full Lennard Jones potential,
\begin{align}
  \phi _{ij}^{WCA} = \left\{ \begin{array}{l}
4\epsilon \left[ {{{\left( {\frac{\sigma }{{{r_{ij}}}}} \right)}^{12}} - {{\left( {\frac{\sigma }{{{r_{ij}}}}} \right)}^6}} \right] + \epsilon \quad {r_{ij}} < r_c \\
0\quad \quad \quad \quad \quad \quad \quad \quad \quad \quad \quad \;\,{r_{ij}} \ge r_c .
\end{array} \right.
  \label{LJ_potential}
\end{align}
The force between pairs of molecules is calculated using the derivative of \eq{LJ_potential} for interactions closer than the cutoff distance,  $r_c = 2.5$. 
The sum of forces on each molecule $i$ is used to obtain the total force.
This total force is then used with Newton's law to obtain the acceleration.
From the acceleration, the evolution of the molecules is obtained by numerical integration, here using the Leapfrog-Verlet scheme.
A timestep of $\Delta t = 0.005$ is employed, with the system initialised as an FCC lattice at a density of $\rho=0.8442$ and temperature of $T=0.722$.
The majorities of the runs are at this statepoint, with a few cases used for comparison in the liquid state at $\rho=0.8$ and $T=1.0$ and solid state with $\rho=1.5$ and $T=1.0$ shown in the appendix section \ref{sec:other_statepoints}.
Periodic boundaries are employed in all three directions.
The simulation domain is a square box of side length $L_x = L_y = L_z = 21.835$ which contains 8788 molecules. 
This was initialised as an FCC lattice at the required density, which is then allowed to melt to a liquid for an equilibration period of 1 million steps.
Having equilibrated the system, the volume averaged pressure is collected inside the domain using a grid of cells.
\begin{table}
\centering
\begin{tabular}{c c c c c c c}
\hline
Case & $N_{\text{cells}}$ & $L_x$ & $\Delta x$ & $t_{\max}$ & $N_{samples}$ & $\mu$ \\
\hline
1 & 46 & 44.64 & 0.97 & 3.0 & 360,000 & 3.35 \\
2 & 32 & 32.25 & 1.01 & 3.0 & 580,000 & 3.12 \\
3 & 22 & 22.87 & 1.04 & 3.0 & 79,000 & 3.36 \\
4 & 36 & 22.46 & 0.62 & 3.0 & 71,000 & 3.37 \\
5 & 147 & 21.98 & 0.15 & 3.0 & 620,000 & 3.23 \\
6 & 281 & 21.83 & 0.08 & 1.0 & 10,000 & 2.69 \\
7 & 281 & 21.83 & 0.08 & 0.5 & 38,000 & 2.68 \\
\hline
\end{tabular}
\caption{Simulation cases with max cells, total domain length, cell resolution, maximum time, number of samples and estimate of viscosity. }
\label{case_table}
\end{table}
This was divided into 7 cells in each dimension and successive layers were cross correlated, which over periodic boundaries allows up to a maximum of 3 layers of cross correlation for a given cell (one self correlation called the zeroth layer and successive layers either side with 3 covering the full 7 cells in the domain).
At 3 cells of cross correlation, averaged over all cell we reclaim the Virial expression exactly.
The correlation in three dimensions quickly becomes prohibitively expensive with number of layers, scaling with $N_{cells}^3$ correlated with and averaged over $N_{cells}^3$ other cells. 
A one-dimensional correlation is shown to perform in a similar way, so is used to explore the effect of increasing number of layers.
The range of cases studied are shown in table \ref{case_table}.
These simulations are used to explore the effect of domain size varying just the $x$ directions keeping $L_y$ and $L_z$ the same, so that $L_x=\{21.83, 30.23, 43.67 \}$ for the various cases studied.
Varying cell sizes were also checked, with $\Delta x = 0.6$ and $\Delta x = 0.15$. Finally $\Delta x = 0.08$ is used for high spatial resolution over short times. 
This allows cross correlations to go up to larger number of cells, going to $147$ (73 layers, $\Delta x = 0.15$) for the full time intervals and some very high resolution $281$ cell runs for short times (140 layers with $\Delta x = 0.08$) to define the short distance and time behaviour.
Statistics collected $N_{samples}$ varied in different cases, as the criteria that the viscosity measured had stopped changing was applied in each case.
Acceleration of the autocorrelation is required for large numbers of cells, aiming to ensure times taken are load balanced between the MD run and autocorrelation calculation time.
The workflow applied ran the MD code Flowmol to collect the grid of VA stresses at each step, with the files written to disk for a set of two temporal correlation times (1200 steps), before using a Python framework to read the entire record and process the space time correlations.
A ram disk was used to accelerate this process, where the file IO operation is not written to disk but stored in a folder which is kept in RAM, allowing fast analysis of the data and preventing disk wear.
At small times the MD is the limiting step, with MD acceleration using MPI applied, but for layers in 1D greater than 147, a GPU accelerated cross correlation is used to roughly match the two simulations.

\subsection{Results and Discussion}
\label{sec:Results}

\begin{figure}
  \includegraphics[width=0.9\textwidth]{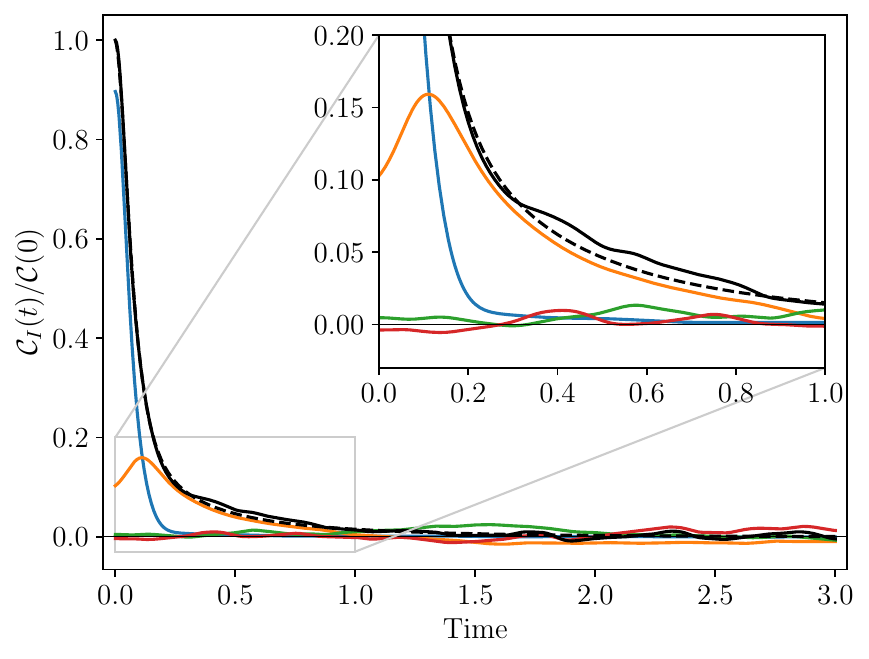}
\caption{$a)$ Plot of contribution to the autocorrelation of stress at each level of cross correlation in a 3D system. 0th layer is the self correlation of the central cell with itself, followed by correlation with the 26 cells surrounding cells in the next later, 64 above and so on, quickly becoming computational prohibitive so only 3 layers are used, giving a total of $7\times 7 \times 7$ cells in the domain as each central cell correlates with 3 either side. Cells are of size $\Delta x =\Delta y=\Delta z \approx 3$ with about $25$ molecules in each. Agreement with the virial (black line) is exact when all cells are cross correlated, and compared to Analytical solution from Eq \ref{Cs} as the dotted black line.}
\label{3DCC}
\end{figure}

We start by looking at the case of a 3D cross correlation, shown in Figure \ref{3DCC}.
The first layer is a shell consisting of 26 cells and so on, as shown in Table \ref{table_layers}.
As this requires the correlation of every cell with every other, this quickly becomes prohibitive so a maximum of 3 layers is considered in the 3D case, requiring $7\times7\times7$ cells in total using a periodic domain. 
\begin{table}[h]
\centering
\begin{tabular}{c|c|c|c}
\textbf{Layer} & \textbf{Cube size} & \textbf{Number of cells in layer} & \textbf{Cumulative} \\
\hline
0 & $1^3$ & 1 & 1 \\
1 &$ 3^3$ & 26 & 27 \\
2 & $5^3$ & 98 & 125 \\
3 & $7^3$ & 218 & 343 \\
\end{tabular}
\caption{Number of cubes in each layer of a $7\times7\times7$ cube}
\label{table_layers}
\end{table}
However, we can see the contribution from the different layers, where the sum of each successive layers $0, 1, 2$ and $3$ in Figure \ref{3DCC} eventually adds up to the Virial when all layers are included.
This is, by construction, the decomposition of \eq{virial_to_VA} into volumes means the sum of the cross correlation of every layer together must be exactly equal to the Green Kubo applied to the Virial.
This total sum of layers which is the autocorrelation of the Virial is shown on Figure \ref{3DCC}  for reference, together with the analytical form \citep{Heyes_et_al_20}.
\begin{align}
C_s(t) =
G\left[
A\,\operatorname{sech}\left(\frac{t}{\tau_1}\right)
+ B\,\exp\left(-\frac{t}{\tau_2}\right)
+ (1-A-B)\,\exp\left(-\frac{t}{\tau_3}\right)
\right],
\label{Cs}
\end{align}
and $G = 24.07$, $A = 0.72637$ and $B = 0.192605$ with timing coefficients $\tau_1 = 0.0492775$, $\tau_2 = 0.182062$ and $\tau_3 = 0.572743$.

A number of interesting observations are possible from Figure \ref{3DCC} , firstly that the self correlation (0th layer) quickly becomes small after a very short time, where the line drops to almost zero well before a time $t=0.5$.
Meanwhile, the interaction with the first shell is growing at the start of the autocorrelation time, reaching a peak before also decreasing to zero well after $t=1$.
The second layer contribution is much smaller and only becomes important much later in the dynamics, adding a peak between the time of 1 and 2.
The susceptibility to noise is apparent in the second later, a trend even more obvious by the third layer where a fluctuating contributions appears only at very late times.
The sum of all layer contributions is exactly equal to the direct calculation for the virial, provided as a useful checksum, and this virial expression gives good agreement with the analytical virial expression.
The cell size is larger than the cutoff radius here, $\Delta x =\Delta y =\Delta z  \approx 3.0$ so we would expect the 0th layer to include most of the first molecular layer for a give molecule.
Each additional shell brings in further molecular layers which increasingly contribute later to the correlation and are less important.

\begin{figure}
 \includegraphics[width=0.98\textwidth]{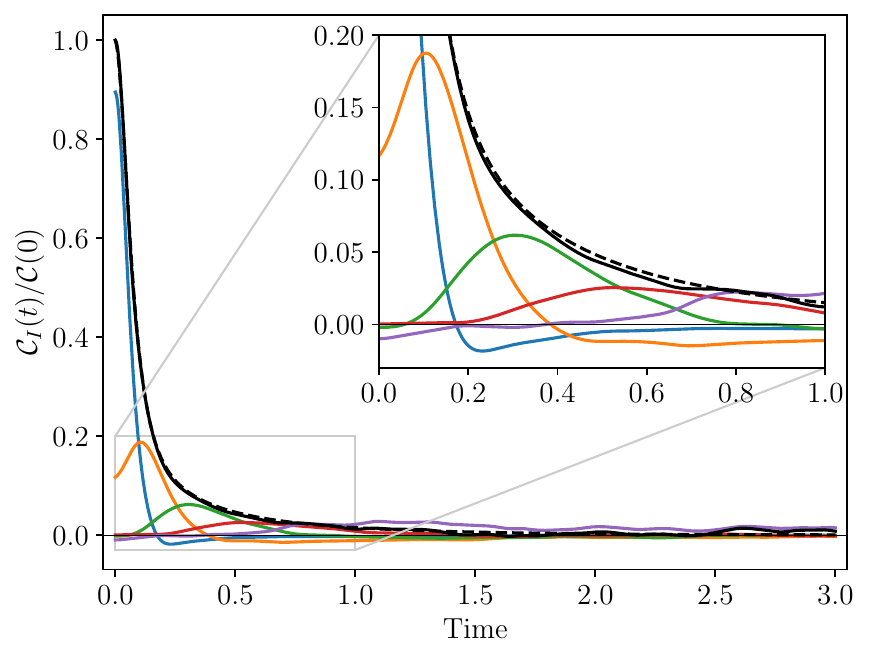}
\caption{Plot of contribution to the autocorrelation of stress at each level of cross correlation in a 1D system. Blue is the 0th layer for the self correlation of the central cell with itself, followed by orange line showing autocorrelation with the cells either side (layer 1), then green (layer 2) the next two layers, then red (layer 3) and purple (layer 4-23 inclusive). Cells are of width $\Delta x \approx 1$ so 3 times the resolution of the 3D case, but with approximately $400$ molecules in each layer. Taking the sum of all layers gives the dotted black line which is exact the virial calculated from the MD system, validated against the analytical form $C_s$ from Eq \ref{Cs} shown by the black line.}
\label{1DCC}
\end{figure}





However, only 3 layers in the 3D case represents 343 cells that need to be cross correlated.
In order to further explore with more layers and smaller bins, as well as look into different domain sizes and orientations, we switch to the 1D case.
This reduces the scaling from $\mathcal{O}(N_{cells}^3)$ to a linear scaling with number of layers in the domain $\mathcal{O}(N_{cells})$.
Using a domain which is long and thin, we can explore the cross correlation in 1D while retaining the essential features of the method. 
This is shown in Figure \ref{1DCC} where 21 layers in $x$ in the same size domain as the 3D case of Figure \ref{3DCC}, with resolutions three time higher $\Delta x \approx 1.0$ but with an order of magnitude more molecules in each layer so better statistics and much cheaper cross correlation.
The shape of the functions are broadly similar to the 3D cases, but the higher resolution allows additional insights, as it is possible to make the layers smaller than the 3D case to explore more granular spatial-temporal contributions.
The 0th layer, $x=\pm 0.5$ assuming we start from the centre of the $\Delta x = 1$ cell.
At this higher resolution now has the intriguing property of becoming negative after the initial period of rapid decorrelation.
This has the character of a reflection, which might be exposing the nature of the atomic structure of the Lennard Jones (LJ) lattice.
An initial stress state decorrelates due to ballistic trajectories before bouncing back due to the molecules hitting their molecular cage.
This negative contribution of the 0th layer also appears in the 1st layer (up to $x=\pm 1.5$) after the same initial growth, peak and decay.
The 2nd layer (up to $x=2.5$) is more pronounced, starting at zero and growing to a peak before also decaying as time increases.
The 3rd layer (up to $x=3.5$) also show these peaks, but each layer shifts this contribution to later times, a trend much more apparent when we look at the integrated contributions of each in Fig \ref{1D_cumulative_perlayer}.
The layers 4 to 23 ($3.5 < x < L_x$) are grouped as they appear to contribute very little to the actual Virial stress over time, except for a noisy longer term peak. 

\begin{figure}
\begin{subfigure}{0.46\textwidth}
\includegraphics[width=\textwidth]{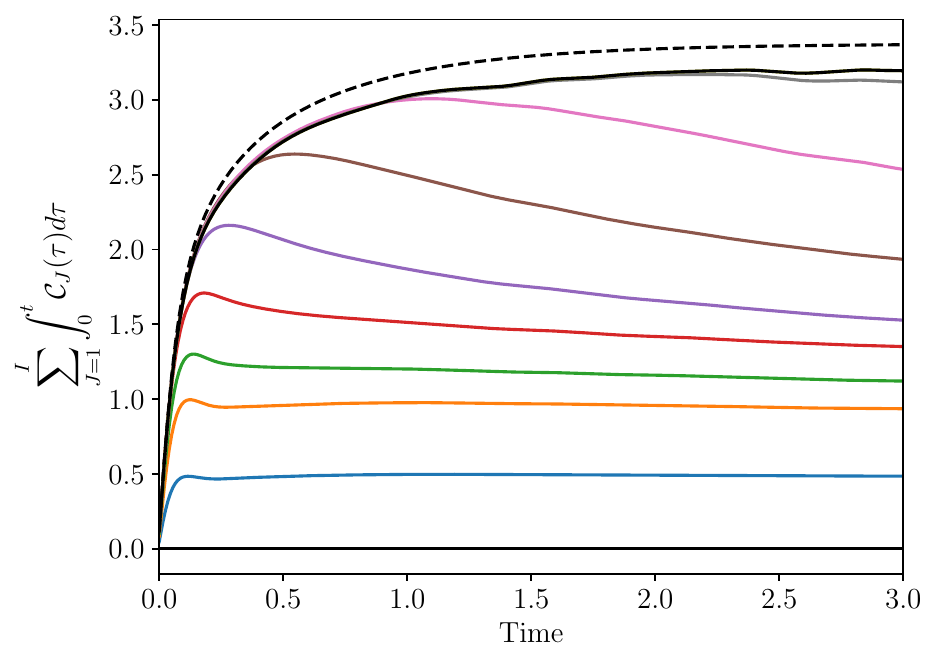}
\caption{Integrated contribution to autocorrelation in time adding varying levels $I$ of cross correlation in 1D case, with sum over $I= 0,1,2,4,8,16,32,64,73$) layers coloured respectively blue, orange, green, red, purple, brown, pink and grey for $64$ with the whole domain $73$ (virial) as black. Integral $C_s$ Eq \ref{Cs} is a dotted black line.}
\label{DCC_cumulative}
\end{subfigure}
\;\;
\begin{subfigure}{0.43\textwidth}
 \includegraphics[width=\textwidth]{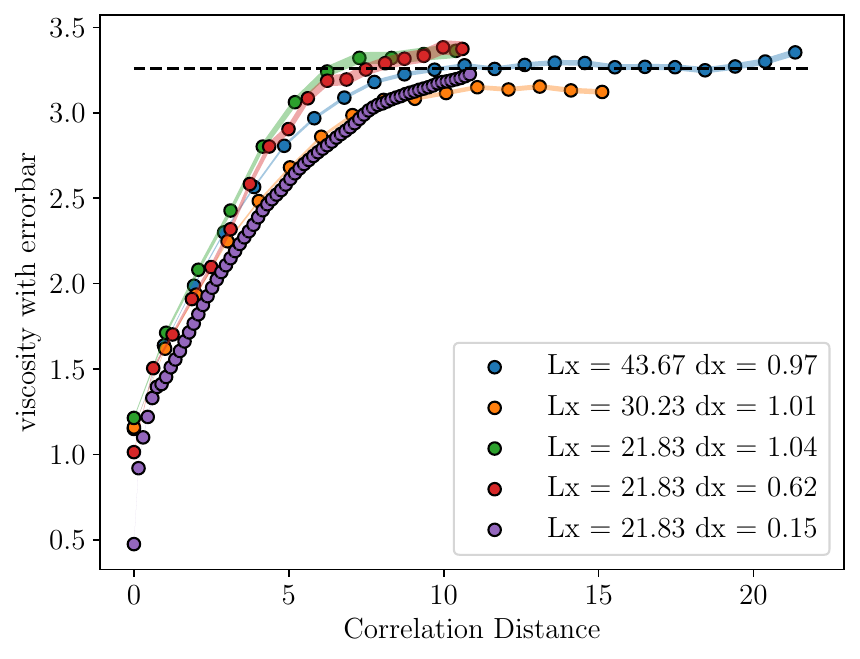}
\caption{The change to time integrated viscosity when including additional layers. Transparent points show viscosity measured for every ensemble, which act like errorbars for the ensemble average line (including black edged circles) obtained from the average of all runs.\vspace{0.24in}}  
 \label{Layers_vs_visc}
\end{subfigure}
\caption{Space and time cumulative integrated correlations showing the plateau in terms of time and space.}
\end{figure}

However, when looking at the cumulative integral  which is shown in Figure \ref{DCC_cumulative}, these contributions are seen to become important at longer times.
A much higher resolution is used in this case, taking the system with 147 cells and $Delta x =0.15$ to explore the importance of spatial contributions.
Each successive layer for $I=1$ to $I=4$ adds a diminishing contribution that seems to apply contributions evenly at all times, with a slight peak at short times.
Beyond $I=4$ to $I=32$, the contributions tend to be increase the cumulative correlation in the short times, and the long time contributions actually seem to be lower.
The layers added from $32$ to $64$ seem to fill in the curve for values when $t>1$. 
One clear trend is the contributions to the viscosity of successive layers become important at later times, for example with the pink line in Figure \ref{DCC_cumulative} we see agreement with the full virial until $t=1$ then the long time contributions are missing.
It is the layers from 32 to 64 that contribute this, and the agreement with virial is seen when we have  around 64 cells.
The cellsizes in this case $\Delta x=0.15$ means 32 cells is a distance of $4.8$ reduced units and 64 cells is approximately $9.6$.
The full expression for all 73 layers is seen to tend to $\mu \approx 3.2$, which under predicts the analytical form from \citet{Heyes_et_al20}, plotted as the integral of the fitting of Eq \ref{Cs} which tends to a value of around $\mu = 3.35$. 
The chosen statepoint is tricky to get consistent viscosity measurements, an independent Green Kubo on much longer runs for this system got a value of $3.2$ which is consistent with result for the full LJ potential reported in past work, D. Levesque and L. Verlet  Mol. Phys. 61 143 (1987); Table 4.

To understand the dependence of physical distance of the integral, the cumulative viscosity in space is summarised in Figure \ref{Layers_vs_visc} which shows how each additional layer adds to the measure viscosity.
A large number of ensembles are run to get good statistics.
Two different domain sizes and two different cell sizes are compared.
Note that the correlations goes outwards from a central cell so the largest correlation in a domain of size $L_x=21.8$ is $L_x/2=10.9$.
Varying the cell size does not effect the resulting viscosity and it appears from Fig \ref{Layers_vs_visc} that a correlation disttance of ~8 reduced units is required for the viscosity to plateau (approximatly 10 cells with $dx=1$ or 17 cells with $dx=0.6$). The resulting viscosity appears to be a value of about 3.2, consistent with literature [D. Levesque and L. Verlet  Mol. Phys. 61 143 (1987); Table 4].
As the same intermolecular forces are simply averaged in different ways when changing box size, the similarity is expected.
To investigate the impact of finite domain size effects, a large domain of 43.67 was tested.
Interestingly, the viscosity is slightly lower than the 21.8 case with correlation over the whole domain, and only increases to a value of 3.2 in the last few cells near the periodic boundaries (around a correlation distance of 22).
An intermediate case with domain length of 30.23 was tested to see if this gave a different value but this provided a notably lower viscosity from the samples.
The standard error is shown as the colour band in Fig \ref{Layers_vs_visc}, which in most cases is smaller than the symbols, suggesting it is not a statistical resolution issue. 
One possible reason for this difference is that at this state points the system is prone to becoming stuck in regions of stable phase space.
This is a common problem with this $T=0.7442$ and $\rho =0.8442$ statepoint, as observed from the work of Woodcock [REF HERE]

We are also in a position to provide some insight into finite size effects here from Figure \ref{Layers_vs_visc}, showing the correlation needs about 10 reduced units to converge in both direction putting the lower limit on domain size to avoid finite size effects at around the box size of 20 used here.
This spatial integration approach is well established in the continuum fluid dynamics literature, where spatial correlations are used to ensure the domain size is large enough to encompass the requisite turbulent structures \citep{Jimenez_Moin_91}.

\begin{figure}
 \includegraphics[width=0.98\textwidth]{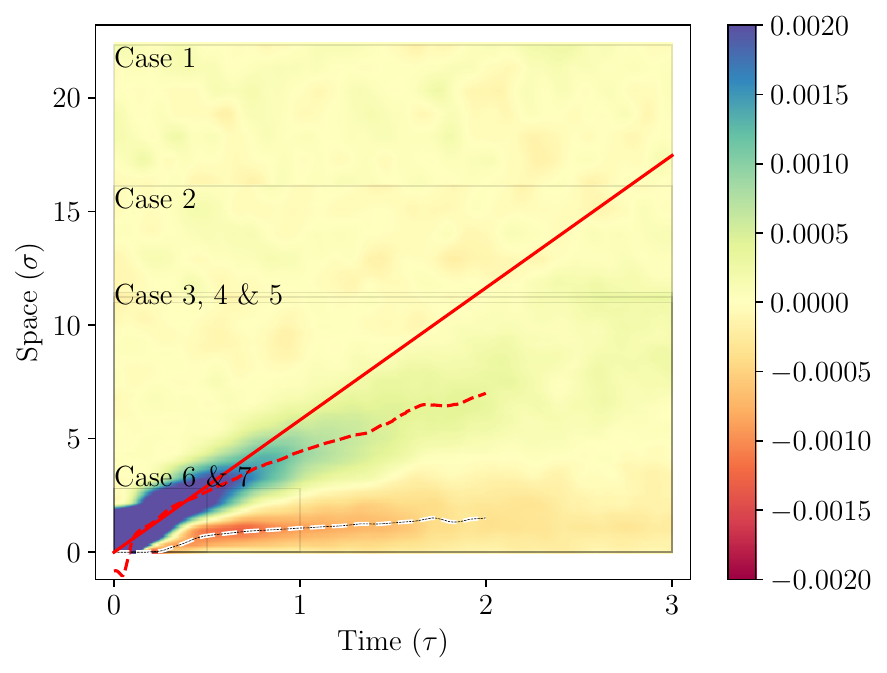}
 \caption{A single spatial temporal contour plot of $\mathcal{C}(r,t)/\mathcal{C}(0,0)$ made by stitching together all the cases run (as outlined in table \ref{case_table}), where boxes show the spatial limits of each case. All cases were interpolated to the highest resolution case to fit them together. This shows the spatial temporal response averaged over all systems studied, with all three domain sizes $L_x=\{43.67, 30.23, 21.83 \},$ sampled at a bin size of $d x=1.0$, together with higher resolution bins run for $L_x=21.83$ with $dx=1$, $dx=0.6$, $dx=0.15$ and $dx=0.08$. The speed of sound $c_s = 5.4$ is shown as a straight line, and the peak value of the travelling waves are identified by a Gaussian fit in space fitted to each peak at every time, with the value shown as a red and white lines to highlight the wave movement. 
}  
 \label{Contour_plot_all_cases}
 \end{figure}

As we increase the number of layers, the noise in our measurement becomes larger.
A natural question is can we truncate in space to improve statistics and get a clearer viscosity measurement.
A similar question was asked in time in the work of Zhang, et al (2015) 
where they use both a fitting to the viscosity as a function of time and a weighting based on the standard deviation of the measurement.
They show as integration time increases, the noise standard deviation of the measurement grows with $std(\mathcal{C}(t)) \propto  \sqrt{t}$.
Similarly, we show the same plot as a function of number of layers in Fig \ref{fit_std}, where the spatial dependence of standard deviation also apparently fits a square root relationship $std(\mathcal{C}(x)) \propto  \sqrt{x}$.
To utilise this and improve statistics, we could assume the space and time dependence of the autocorrelations are independent of each other and approximate the response as $\mathcal{C}(x,t) = \mathcal{C}_x(x) \mathcal{C}_t(t)$.
The resulting functional form that can be obtained making this assumption is included in the appendix to potential spatial truncation strategy.
To test if the assumption of independence is reasonable, we combining all the different domain sizes and resolutions, we present a single temporal-spatial contour plot of the autocorrelation $\mathcal{C}/\mathcal{C}_0$ in Figure \ref{Contour_plot_all_cases}.
The value of contour is normalised by the initial value $\mathcal{C}_0=\mathcal{C}(x=0, t=0)$ so therefore the contour has a value of one at the origin.
From Figure \ref{Contour_plot_all_cases}, it is quite clear the two dimensions are strongly correlated and cannot be simply decomposed, i.e. $\mathcal{C}(x,t) \ne \mathcal{C}_x(x) \mathcal{C}_t(t)$ and a non-linear function in $x$ and $t$ will be required to approximate the decorrelation.
In order to do this, we start by trying to understand the form of this spatial temporal response.


This master plot shows the response of a Lennard Jones liquid to an instantaneous state of stress and how this decorrelates as a function of both space and time around an arbitrary point in space.
This is a  spatial-temporal extension of the Green Kubo function.
The space integral of this contour plot would gives the exact time curve of the normal form of Green Kubo.
A further integral in time would give the viscosity coefficient.
The use of varying bin sizes with $\Delta x \to 0$ ensured the shape of these features are not an artifact of different averaging binsizes and the different domain sizes ensure the presented results are system size effects independent.
The strongly positive blue region is a wave of shear pressure which starts at the origin (maximum correlation) and moves outwards with a speed that initially matches the speed of sound in this system $c_s = 5.43$.
This definition is consistent with the for the truncated shifted LJ  model considered by \citet{Thol2015} and \citet{Heier2018}, who provides thermodynamic properties, including the speed of sound, over a broad range of temperature and pressure from differentiation of the Helmholtz free energy.
Note that interactions between molecules means the influence of an atom travels ahead of the wave centre at time zero, seen as a blue region spreading a distance equal to the cutoff length $r_c = 2.5$ initially. 
However, no influence of the stress  can travel faster than the speed of sound, so the top half of Figure \ref{Contour_plot_all_cases}, above the line given by the speed of sound plus cutoff length $c_s t + r_c$, can be assumed to be purely noise.
This suggests an immediate use of this approach, by truncating averaged samples to spatial-temporal results below the dotted red line (speed of sound + cutoff), statistical noise in the Green Kubo can be reduced.
However, a much more useful approach is possible as the data has a very clear mathematical structure.
The centre of this wave region is tracked by fitting a Gaussian and it can be seen that after a time of  $t\approx 0.5$, the speed drops below the speed of sound.
This is attributd to the wave moving through a deforming liquid and the diffusion of the wave packet slowing the speed.
The speed of this wave follows a $\sqrt{t}$ form which is $\approx t$ at low times.
Another interesting feature of the liquids spatial temporal stress response is a negative peak which appears to follow the main positive one. 
This peak is likely the first layer of the liquid bouncing back after the initial wave has passed through, a well documents feature of the liquid cage.
Recall the integral of this contour plot over both space and time is the viscosity.
This negative region represents an anti-peak that has negative contribution to viscosity (an anti-viscosity).
While the main wave pushes through the fluid propagating a state of positive shear stress correlation, the negative region appears to stay centred on the first layer of the liquid.
Both peaks decrease and spread out in ways that have deep connections to the structure of the liquid state.



In order to understand the nature of the wave front, we fit a Gaussian function at each time to both waves. 
Combining both, this is very well approximated by the sum of two Gaussians, a positive (leading wave) $c_1$ and negative (rebound) $c_2$,
\begin{align}
\mathcal{C}(x,t)=
\underbrace{A_1(t)\exp\left[-\frac{(x-x_1(t))^2}{2\sigma_1(t)^2}\right]}_{c_1}
+
\underbrace{A_2(t)\exp\left[-\frac{(x-x_2(t))^2}{2\sigma_2(t)^2}\right]}_{c_2}
\label{2gaussian_fn}
\end{align}
\begin{figure}
\includegraphics[width=0.8\textwidth]{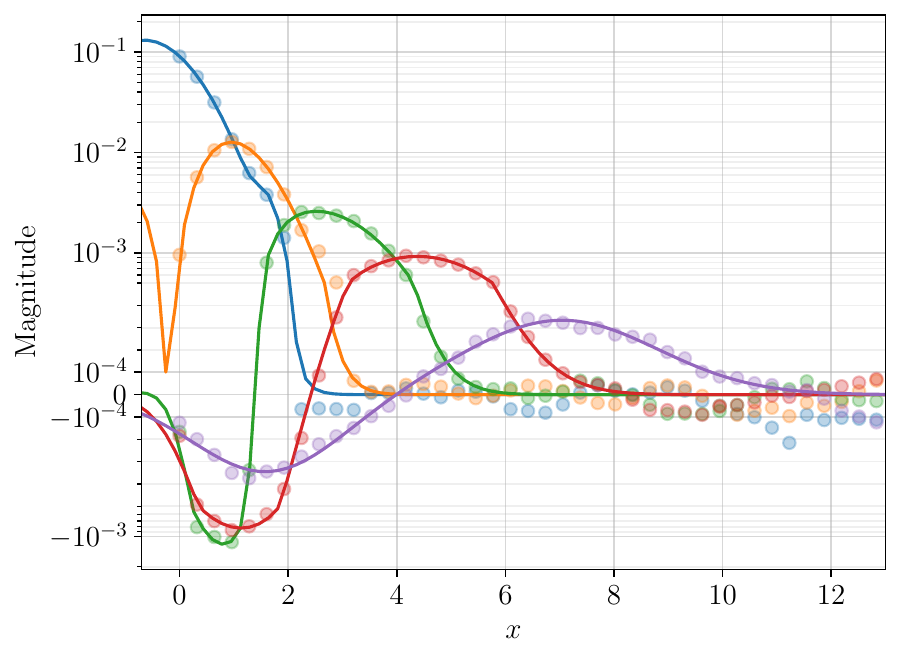}
\caption{Fitting a double Gaussian from Eq \ref{2gaussian_fn} to the spatial signal at a range of times, log scales are used above $|10^{-4}|$ with a linear scale near the origin. The times shown are $t=[ 0, 0.15, 0.5, 1, 2]$.}
\label{fitting_fronts}
\end{figure}
The least-squares fitting of Eq. \ref{2gaussian_fn} at every timestep is shown in Figure \ref{fitting_fronts}.
The fitting process uses a form of curve peeling, where the positive Gaussian is first fitted to a subset of the spatial data within $w_{\rm size}$ of the estimated peak position, $|x-x_1^{\rm guess}|<w_{\rm size}$, with $w_{\rm size}=8$ found to work well.
As an initial guess, the peak is assumed to be the $x$ location of the maximum value,
$x_1^{\mathrm{guess}} = \arg\max_x \mathcal{C}(x,t)$
over the spatial domain.
The fit is then weighted so that points near this assumed peak are more strongly fitted than points further away, which was found to give a better fit as outlying points were more susceptible to noise. The one-dimensional uncertainty weighting of SciPy's (v1.18) \texttt{curve\_fit} is used, with
$\sigma = 1/[1+\beta-1)c_1^{\rm guess}]$ where $c_1^{\rm guess}$ is the Gaussian $c_1$ evaluated with $x_1=x_1^{\rm guess}$ and $\sigma_1=w_{\rm weight}=w_{\rm size}/2=4$. The weighting is evaluated at each of the discrete spatial data points, with $\beta=10$.
The maximum value of $\mathcal{C}$ and position are used as the initial guesses in the curve fit with a starting $\sigma_1$ value taken as unity.

It is worth noting that at short times the wave has not travelled far enough to expose the full Gaussian shape, with only one side apparent.
Despite this, it still seems to be well fitted with a negative $x_1$ position so only one side of the Gaussian is being fitted to the data, shown at the first time in Figure \ref{fitting_fronts}. 
Once the full profile is apparent in the data, the fit appears to be a very good approximation of the travelling wave in the fluid. 
The left side on the fitted positive peak $c_1$ near zero is then used to determine the upper bound of subset of $\mathcal{C}$ values to fit the negative Guassian $c_2$ to. 
The lower bound is $x=0$ and the x location of the minimum $\mathcal{C}$ value in this range is used to obtain the guess for the negative Gaussian centre $x_2^{guess}$.
No weighting is used here as the negative peak tends to be well defined with low amounts of noise.
This fitted negative peak is measuring the negative correlation following the leading wave.
This is attributed to a rebound in the molecular cage remains located at the second layer from the centre of the correlation. 
The peaks can be seen to be moving over time, stretching out and decreasing significantly in magnitude, which is why a log scale is required.

The Gaussian peaks are approximating the variation of the autocorrelation in space at every time.
The data has 600 timesteps of $\Delta t =0.005$ so $0<t<3$ and the two Gaussian peaks could be fitted in space at every timestep. 
In practice after around 400 timesteps, when $t=2$, the noise makes the fitting of the positive peak impractical with maximum values no longer reliably being the peak of the wave and instead thermal noise in the fluid.
The exception is the negative peak which still shows good signal to noise ratio and so could be fitted over the entire autocorrelation duration.
Using these fits at each time, the time variation of the coefficients, $x_1$, $x_2$, $\sigma_1$, $\sigma_2$, $A_1$ and $A_2$ can be explored. 
If these have a particular pattern, this exposes insights into the fluid itself as well as potentially suggesting a route to modelling the wave propagation in the fluid and ultimately the autocorrelation of a molecular fluid itself.
To this end, we plot the various coefficients and apply fits.
 \begin{figure}
 \includegraphics[width=0.97\textwidth]{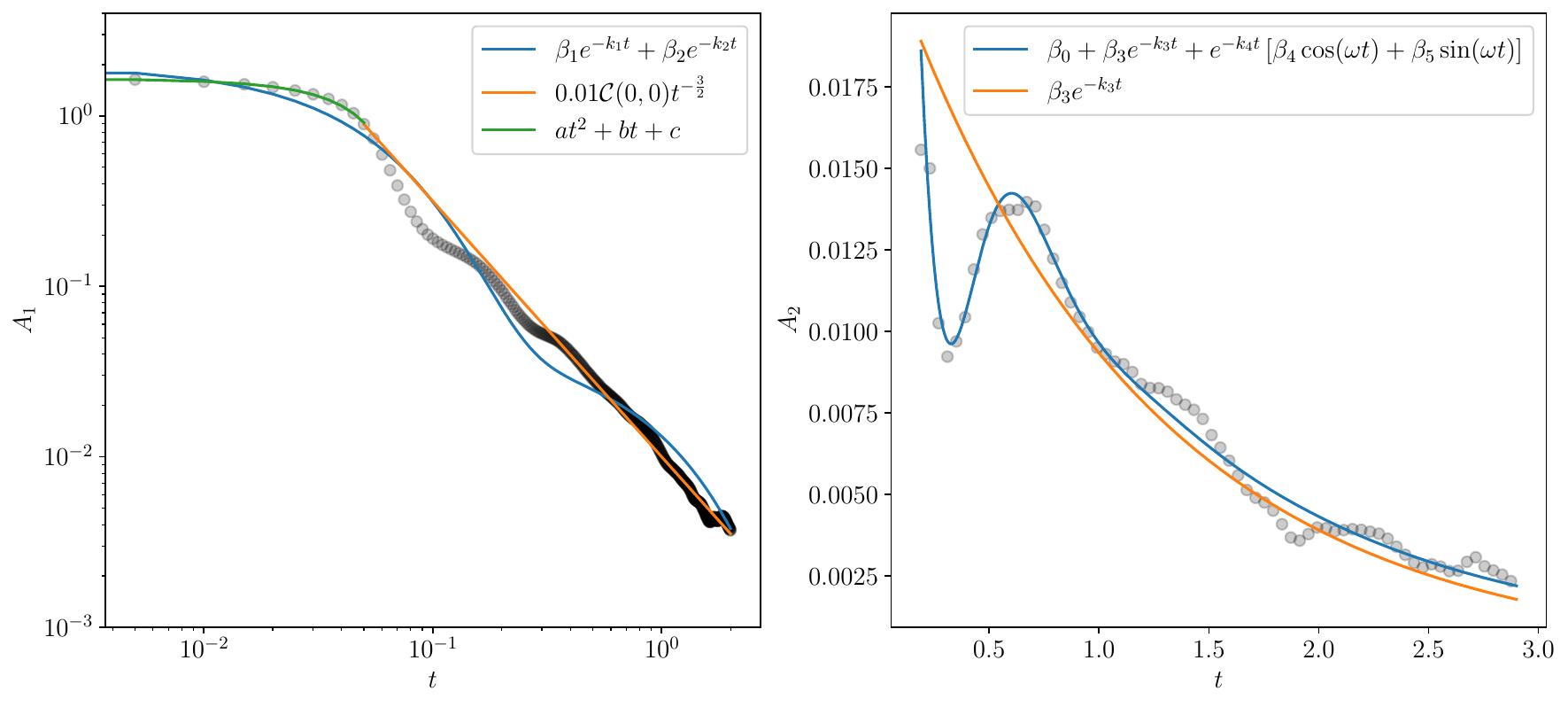}
\put(-420,30){$a)$}
\put(-190,30){$b)$}
 \caption{Coefficients of exponential terms with fitting to the normal pair of exponential used in autocorrelations to match short and long time behaviour. Interestingly, a direct fit to a $t^{-3/2}$, matching the famous Alder (1970) work, and an improvement to the fit for the negative peak that includes the vibration nature of the wave moving between atoms in $A_2$. }  
 \label{A1_A2_fits}
 \end{figure}
 The magnitude of the Gaussian peaks as a function of time is shown in Figure \ref{A1_A2_fits}.
Figure \ref{A1_A2_fits}a) shows the height of the positive peak $A_1$ and an initial guess for a time dependent fit is the same double exponential fit used for the total viscosity.
\begin{align}
\beta_1 e^{-k_1 t} + \beta_2 e^{-k_2 t}
\end{align}
with $\beta_1 = 0.00415$, $k_1 = 1.25$, $\beta_2 = 0.173$ and $k_2 = 19.5$.
This can be seen to generally give a reasonable fit at short and long times but with departure in the intermediate time.
However, it is seem that a much better fit, especially to the long-time behaviour, is given by a power-law decay of the form
\begin{align}
A_1(t) = a \mathcal{C}(0,0) t^{p} + c,
\label{A_fits}
\end{align}
where \(\mathcal{C}(0,0)\) denotes the initial value of the correlation function. In this case, \(\mathcal{C}(0,0)=0.0903\). The fit was applied for \(t \geq 0.190\), corresponding to timestep \(N_{t}=10\). 
The parameters of the power-law are \(p=-1.50\), \(a=0.01\,\mathcal{C}(0,0)\), and \(c=0\).
The least square fit gave a slightly different set of values, with $a = 0.008$; $c = 0.0014351$ and $p=-1.48$. 
However, the value of $p$ is very close to $-3/2$ and the coefficients are rounded to give a concise form.
The fact that the profile on a log-log plot is linear in the long time, with a gradient that agrees well with $t^{-3/2}$, is a very exciting result.
This agrees with the classic findings of Alder and Wainright (1970) who show the long time table in velocity autocorrelation has the form $t^{-D/2}$ with dimension $D=3$, which has a link to larger-scale fluid dynamics.
As time tends to zero, the log plot diverges to infinity so a short time could be approximated by a polynomial, here $-18.24t^2 -0.438t + 0.150$.

For the magnitude of the negative peak in Figure \ref{A1_A1_fits}$b)$, the oscillatory nature of the molecular structure is apparent, so a decaying double exponential with a sinusoidal component seems a reasonable fit.
\begin{align}
A_2(t) = \beta_0+\beta_3e^{-k_3t}
+e^{-k_4t}\left[\beta_4\cos(\omega t)+\beta_5\sin(\omega t)\right].
\end{align}
The fitted parameters were \(\beta_0=3.75\times10^{-5}\), \(\beta_3=2.02\times10^{-3}\), \(k_3=0.871\), \(\beta_4=4.23\times10^{-3}\), \(\beta_5=-1.09\times10^{-4}\), \(k_4=5.33\), and \(\omega=8.34\).
However, taking just the exponential decay peak seems to broadly fit the overall trend (orange line), which is a simpler model for this component,
\begin{align}
A_2(t) = \beta_3e^{-k_3t}.
\end{align}

\begin{figure}
 \includegraphics[width=0.97\textwidth]{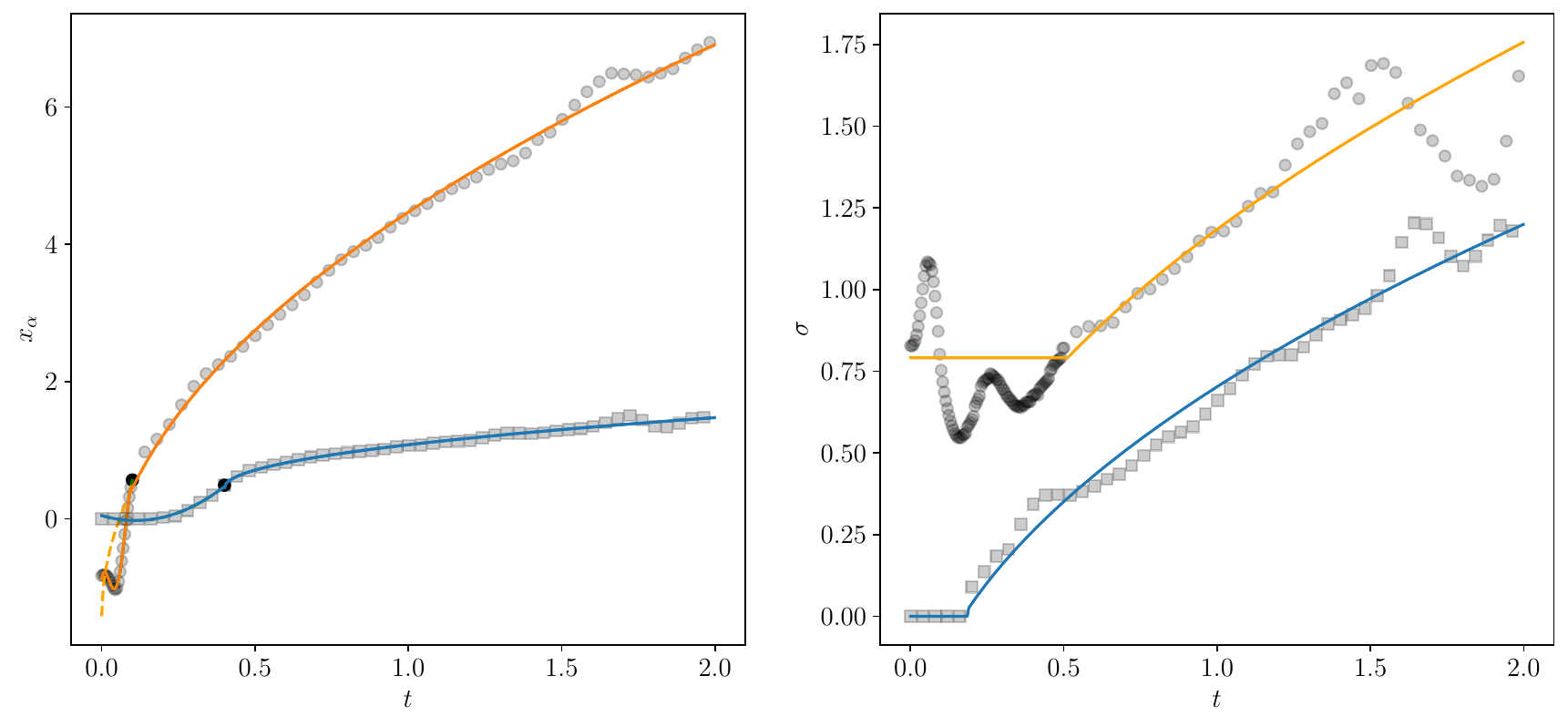}
\put(-430,190){$a)$}
\put(-190,190){$b)$}
 \caption{Fitting of the centre of the two Gaussians $x_{alpha}$ and their standard deviation $\sigma_{alpha}$. The main fits in all cases are square root functions with either constant values below the regions where th Gaussian is fully formed or power law fits.}  
 \label{x_sigma_fits}
 \end{figure}
\begin{figure}
\begin{subfigure}{0.44\textwidth}
\includegraphics[width=\textwidth]{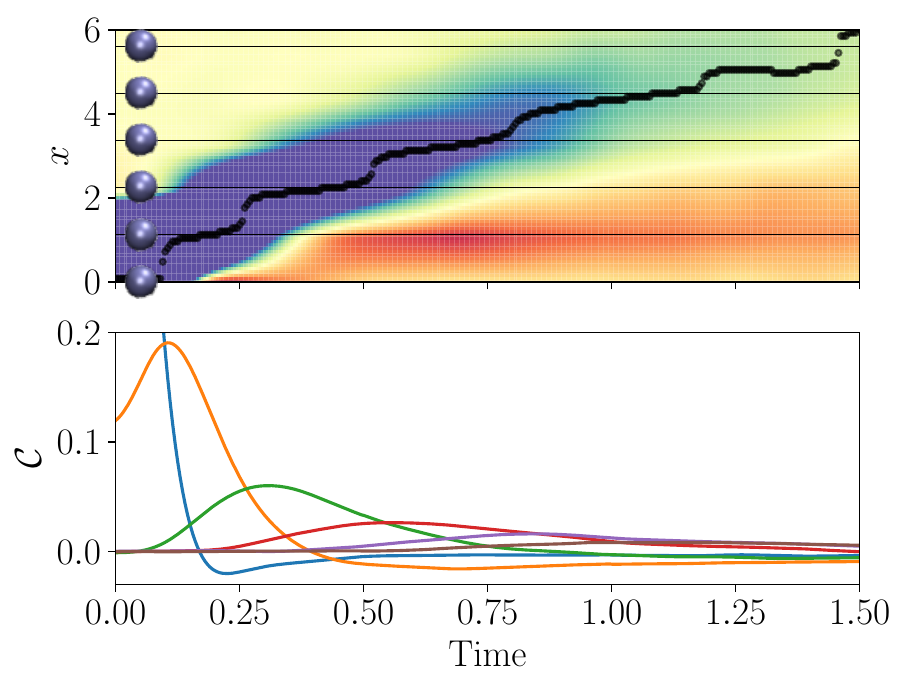}
\caption{Lines at a spacing of $2^{1/6}$ indicate molecular location overlaying colormap, with the location of maximum value shown as black points. Time correlations along lines shown below. The blue line is the $x=0$, orange is $x=2^{1/6}$, green $x=2\times 2^{1/6}$, red $x=3\times 2^{1/6}$, purple $x=4\times 2^{1/6}$ and brown $x=5\times 2^{1/6}$}
\label{216th_spacing}
\end{subfigure}
\;\;
\begin{subfigure}{0.44\textwidth}
 \includegraphics[width=\textwidth]{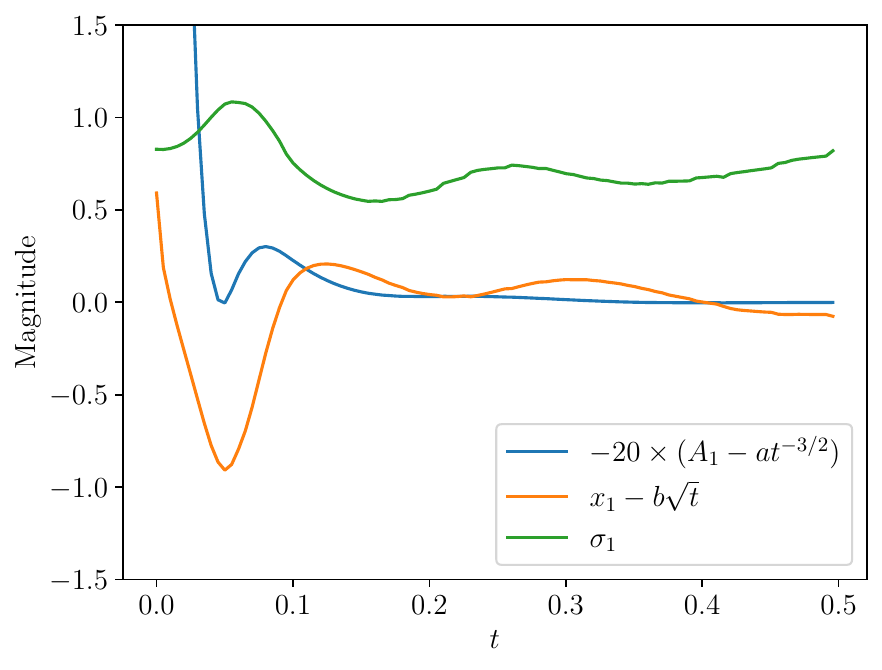}
\caption{The plot of the difference between coefficients fitted each time and the various functional forms used to fit these coefficients. Magnitudes are scales by a factor to show on the same scale. This highlight the molecular structure and how the limitations of a smooth fit.}  
 \label{short_time_oscillations}
\end{subfigure}
\caption{The molecular structure of the fluid is not expected to be fitted by a continuum style Gaussian model of a wave, and the effect of the structure is highlighted in these figures.}
\end{figure}

For the centre of the waves $x_1$ and $x_2$ in figure \ref{x_sigma_fits}a) and widths  \(\sigma_1\), and \(\sigma_2\) in figure \ref{x_sigma_fits}b), they are well fitted by a square root form. 
For \(x_1\), the fit was performed for \(t\geq0.100\) and gave
\begin{align}
x_1(t) & = x_1^i + m_1\sqrt{t}, \nonumber \\
x_2(t) & = x_2^i + m_2\sqrt{t-t_2^i},
\end{align}
with \(m_1=5.89\) and \(x_1^i=-1.42\). For \(x_2\), the fit was performed from \(t=0.401\), and \(m_2=0.805\), \(x_2^i=0.455\), and \(t_2^i=0.401\) using the time shifted form.
The speed of the front is based on a shift square root term, with the $x_2$ shifted in time because the full peak does not exist until the wave has propagated past the first atomic later, around time $t=0.16$.
The corresponding square-root fits for the widths were
\begin{align}
\sigma_1(t) & = \sigma_1^i + m_{\sigma_1}\sqrt{t}, \nonumber \\
\sigma_2(t) & = \sigma_2^i + m_{\sigma_2}\sqrt{t}, 
\end{align}
with \(m_{\sigma_1}=1.38\) and \(\sigma_1^i=-0.199\), fitted over \(0.501\leq t\leq1.25\), and \(m_{\sigma_2}=1.20\) and \(\sigma_2^i=-0.497\), fitted for \(t\geq0.185\).
Again the short time behaviour does not follow this trend due to oscillations and difficulty fitting when only half the Gaussian is present.
The coefficient of the spreading rate is very close in the two fits, suggesting this is a fluid property likely related to diffusion or viscosity itself.
The fitting of Eqs \ref{x_fits} and \ref{sigma_fits} are shown on Figure \ref{x_sigma_fits}, the choice of starting point $x_{\alpha}^i$ or $\sigma_{\alpha}^i$ is chosen by hand, often iterating to ensure the remaining function gives a good fit.
The fit before this is shown as a quintic on the $x_1(t)$ graph and a cubic on the $x_2(t)$ graph, although these regions are already a poor fit due to the molecular structure as discussed next.
For the negative peak, no part of it exists until after $t>0.1$ so cannot be fitted, so again a polynomial could be used along with a shifted square root function to when it is possible to fit.
The oscillatory nature of the width of the distributions $\sigma_1$ components are very apparent in Figure \ref{x_sigma_fits}b).
As the wave front moves from one atomic position to the next, both the location of the peak and the apparent width demonstrate a jump. 
This appears in the fitting coefficients as oscillations which have a passing resemblance to the radial distribution function (RDF) \citep{Rapaport}, especially in the $x_1$ component which essentially passes over successive layers starting from a initial bin which likely has an atom in.
The VA stresses are necessarily taken for arbitrary bins in space so are not atom centric like the RDF, so this effect is somewhat smeared.
The various layers of the fluid are highlighted in Figure \ref{216th_spacing} where lines are drawn at intervals of the energy minimum in a LJ fluid $x=2^{1/6}$.
The maximum value of the correlation function is shown in Figure \ref{216th_spacing} as a set of black crosses which highlight the nature of the wave propagation moving from atom to atom by a roughly flat region inside a layer followed by a jump as it moves to the next layer.

The Figure \ref{216th_spacing} bottom shows the decorrelation of each atomic later below, the value of the contour in time plotted along the successive black lines.
This clearly shows the peak as it moves from one atom and is passed to the next one by correlation and the subsequent decay of this state of stress in the layer once it has reached a peak.
The impact of this molecular structure is somewhat smoothed by the continuous nature of the interaction between molecules, where a wave isn't simply a hard sphere collision but a gradual increase in the force of one atom on the next as the shear wave moves through.
Recall we are measuring the state of stress $P_{xy}$ in a box and how it is passed through space to the adjacent atoms.
Given the timescales and the dense nature of the fluid, the majority of this is expected to be a configurational part, so $P_{xy} = \sum f_{xij} r_{yij} \ell_{ij}$ which is intermolecular force in the direction of the wave propogation $x$ multiplied by separation in the direction orthogonal to this motion $y$ (or vice versa, separation in the direction of propgation and force in $x$).
It is imagined the state of stress at a central atom interacts through this force to move an adjacent atom layer, increase the stress in this next atoms which in turn pushes it into the next layer.
At short times before the liquid has time to deform substantially, this is expected to be entirely a wave based mechanism so the structure appears more obviously, see in Fig \ref{short_time_oscillations}.



%

The neat form of the two Gaussian suggests a Green theorem solution to an advection diffusion equation which might give physical interpretations to the mathematical forms of these fits.
This all assumes a continuum style liquid, but Figure \ref{216th_spacing} top highlights the limitations of this continuum picture.
The wave moves through a liquid, with quick propagation through the interaction between the molecules followed by slow movement as the molecules needs to physically move before it can interact the next one. 
Gradually the spacing seems to deform, but for the first 1 time unit and 5 layers it appears to stay relatively well structured with the 6 lines spaced by $2^{1/6}$ broadly corresponding to the same physics at each place.
Also of interest is the line at zero and the first layer ($2^{1/6}$) have a very clear negative correlation region, which is most apparent along the first layer.
This represents the anticorrelation due to the stress state at the origin causing the first atom next to it to carry the wave, then rebound in its molecular cage in a way that is anti-correlated with the original signal.
This can be pictured as a state of stress, starting at the time space origin, rippling out through all the surrounding atoms.
The positive peak sits throughout the simulation on the first layer, suggesting this first layer acts as a molecular cage providing a lower viscosity by slowing or reversing the decorrelation of the central atom.

 \begin{figure}
 \includegraphics[width=0.8\textwidth]{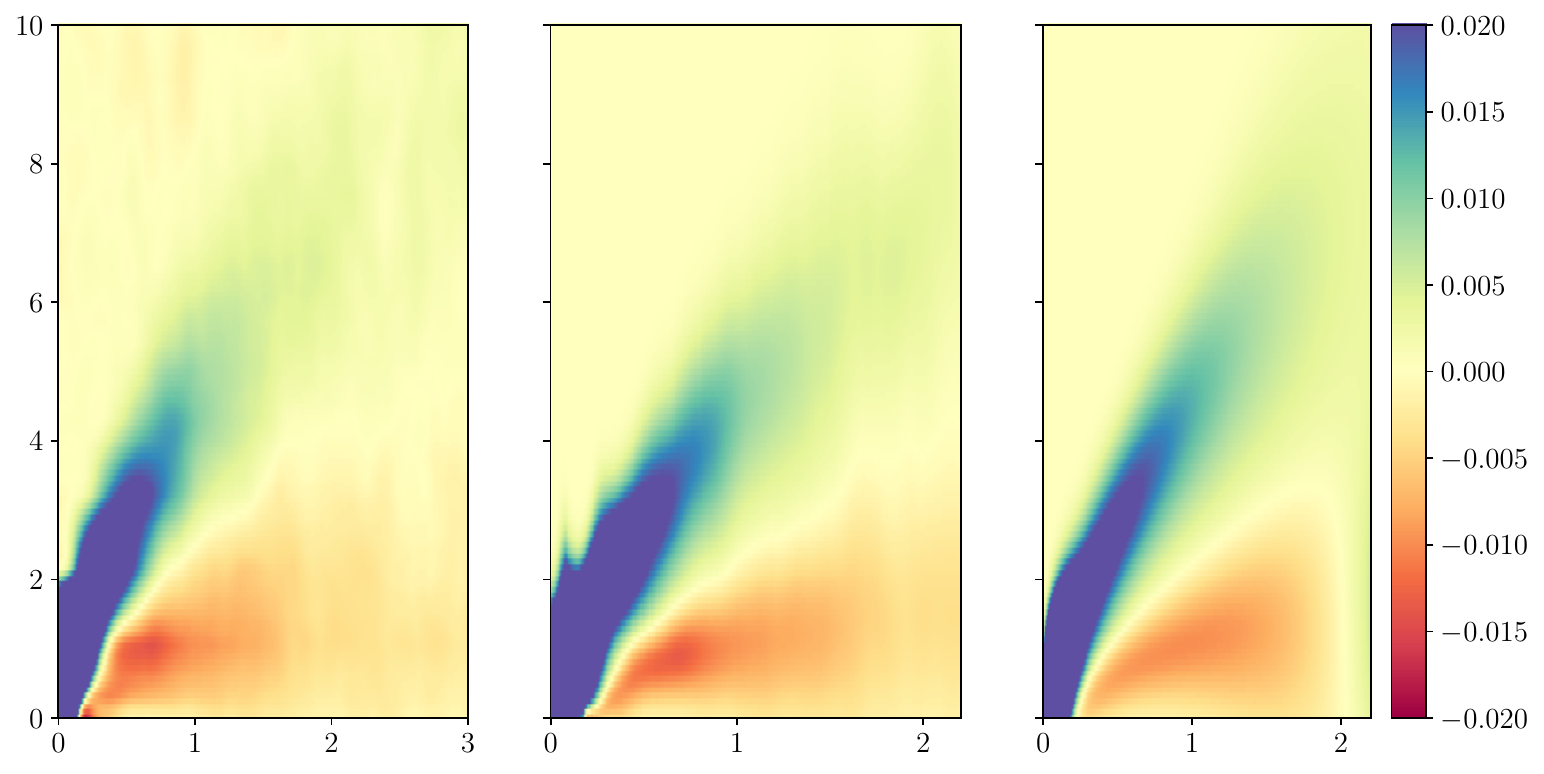}
\put(-355,165){$a)$}
\put(-235,165){$b)$}
\put(-115,165){$c)$}
 \caption{The fitted surfaces, with original data (left), fit every timestep (middle) and fit using functional forms (right)}.  
 \label{reconstructed_waves}
 \end{figure}
\begin{figure}
 \includegraphics[width=0.8\textwidth]{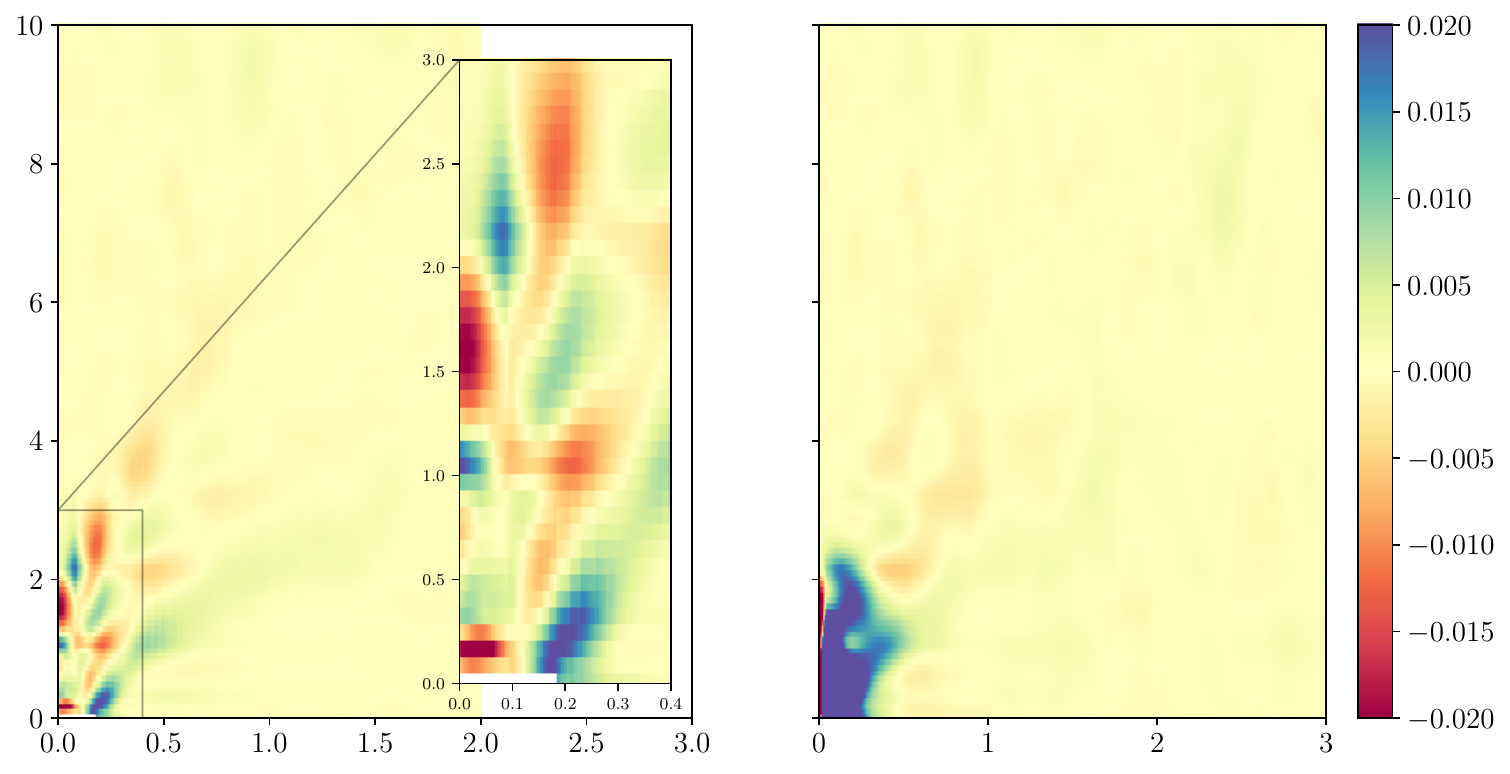}
\put(-350,170){$a)$}
\put(-160,170){$b)$}
 \caption{Errors in reconstructed waves with the model refitted at each timestep (left) and the model with all parameters using close form expressions (right) }.  
 \label{errors_in_waves}
 \end{figure}

Bringing together all the fitting in this section, the original spatial-temporal function is shown in Figure \ref{reconstructed_waves}$a)$ compared to the two Gaussians fitted at each time in \ref{reconstructed_waves}$b)$ and the form predicted using the close form fits of the coefficients Eqs \ref{A_fits}, \ref{x_fits} and \ref{sigma_fits} giving the final fitted form,
\begin{align}
\mathcal{C}(x,t)=
 0.01 \mathcal{C}(0,0) t^{-3/2}\exp\left[-\frac{(x+x_1^i-\sqrt{t})^2}{2 (\sigma_1^i + m_{\sigma_1}\sqrt{t})^2}\right]
+
\beta_3 e^{k_3 t} \exp\left[-\frac{(x+x_2^i-\sqrt{t-t_2^i})^2}{2(\sigma_2^i + m_{\sigma_2}\sqrt{t})^2}\right]
\label{Fully_fitted}
\end{align}
in Figure \ref{reconstructed_waves}$c)$.
The approximation of the function in terms of the Gaussian can be seen to work very well.
The error between the MD results and the fitting every time is shown in Figure \ref{errors_in_waves}$a)$.
Subtracting the two Gaussian wave equation fitted every time from the MD solution, provides a deep insight into the molecular structure in Fig \ref{errors_in_waves}$a)$ with a zoom shown in the insert. A series of peaks and troughs are exposed as the wave sweeps through the material passing from atom to atom.
This is not possible to see in the original contour plots of Figure \ref{reconstructed_waves}$a)$ and only by removing the overall trend of the continuum form of the moving waves is this, uniquely molecular contribution, left.
Each peak or trough is likely showing the location of a molecule and how it bounces from positive to negative.
As the wave moves out, these peaks also appear to stretch as the structure deforms from the initial structure in the liquid.
The complexity of this lattice vibration also has important implications in the fitted forms based on closed form relations of the coefficients of Figure \ref{reconstructed_waves}$c)$.
It is unlikely a functional form for $x$ or $\sigma$ can capture this complexity to give a model for the lattice like contours of Fig \ref{errors_in_waves}$a)$
As a result, there are large errors observed between the MD data and the fully fitted form of Eq \ref{Fully_fitted}, as highlighted by the difference in Figure \ref{errors_in_waves}$b)$.
These errors seem to be localised to $t<0.5$ and $x<2.5$, the point where the molecular nature is most apparently.
These fits can try to be improved by polynomial fittings at short time or even prony fits or sinusoidal corrections, but these are unlikely to generalise and do not yield the deeper physical interpretations that the Gaussian wave packets have.
More importantly, short time and local measurements are the easiest to get from molecular simulation.
Consider that to obtain these in a simulation with $\Delta x = 0.6$ and $\Delta t = 0.005$ we would need to correlate over just 50 timesteps in time and 4 layers in space to capture this essential data. The two Gaussian form can then be fitted to this short time and space data and the remaining time and space predicted using the fitted functional forms which can be integrated analytically.
The integral of this fitted form provides a direct estimate of viscosity without having to run long simulations.
It is likely that more than just $t<0.5$ and $x<2.5$ region would be needed for certainly any given fit has validity to long time and space distance.
This would require careful tests for any new system, systematically increasing time and space measurement to ensure the fitted form are predictive beyond the current fitted data.
For example the fit can be applied on the curve in $t<0.5$ and $x<2.5$, before being used to predict a large spatial and temporal region (say $t<1.0$ and $x<5$) and then the test takes this larger region to ensure the fit is reasonable. 
The region could then be systematically increased to obtain a plot of change from previous estimate vs. $t_{max}$ and $x_{max}$, until a small enough value of change is obtain to suggest convergence.
For the small $mathcal{O} (9000)$ Lennard Jones atoms here, this is already potentially a considerable time saving in the required simulations, but for complex systems when atom numbers can be order many millions or for machine learning potentials which are often an order of magnitude more expensive to simulate, the potential benefit can be significant.
This has the potential to enable viscosity to be determined in simulations which are too large or expensive so this is not be possible without these techniques.
However, it is worth noting that this has been paramterised on only one system and for a relatively simple LJ potential, so it is not clear how well this will generalise.
Some insight is given for the case of a different LJ liquid state point and even a solid in the appendix section \ref{sec:other_statepoints}. 
Although fitting is not applied on these cases it is clear the function have forms that would be amenable to such approaches, albeit with a function to capture the oscillatory components needed in solids.
It is also potentially true that fluids with long-time viscoelastic behaviour would see this occur in the longer time tail, something which would require long temporal spatial data.
This prevents short fittings from being predictive of behaviour that only emerges with long times and distances, a problem shared by Green Kubo.
The collecting of statistics in space and time instead of simply time as in the Green Kubo is relatively cheap and essentially provides more potential statistical power from the same simulation steps. 
The improve statistics with this spatial data is enabled as multiple space and time origins can be used, providing more statistical averaging from the same run times. 
Improve statistics also come from neglecting the region faster than the speed of sound and, in the best case, fitting a function to the data which approximates the majority of the behaviour and can totally removes any noise or uncertainty in the data.  
This fitting function also potentially exposes new insights into the fluid, in this case the spreading rate of the Gaussian wave packet has a clear link to the diffusion in the fluid while the decay in magnitude appears to be linked to how quickly a velocity in the fluid decorrelates with $t^{-3/2}$ form.
These insights can allow a fitted form in the best case that can be used to predict behaviour as time tends to infinity, giving a clearer measure of constant viscosity or exposing novel viscoelastic behaviour.

\section{Conclusions}
\label{sec:Conclusions}

Spatial decomposition of a single global pressure into a local binwise form is possible.
Using this local stress in the Green Kubo stress form allows the definition of a spatial temporal version of Green Kubo based on the cross correlation of every positions and time with every other.
This allows multiple space origins to improve statistics, which is prohibitive in 3D beyond more than a few layers, in 1D it can allow many layers and high resolution and with GPU acceleration this can become comparable to the simulation times.
A detailed parameter study using varying resolution and domain sizes shows that the required spatial distance to ensure converged results is around ten reduced units, and beyond does not contribute to the autocorrelation except in the form of noise. 
This suggests that using a spatial temporal correlation and truncating in space could be used to reduce the required samples while also improving signal to noise ratio.
Plotting the spatial temporal plots of the integrand of the Green Kubo expression identify clear structure exposing a positive travelling wave followed by a negative wake.
This function is almost perfectly fitted with a two Gaussian function which captures the wave nature of the autocorrelation function starting from the origin and moving out initially with the speed of sound before slowing to a square root style dependence.
The width of the peak also spreads out with Gaussian width proportional to the square root of time.
Perhaps most interestingly from a liquid state point of view, the magnitude of the peak decays as $t^{-3/2}$ which is the predicted form of autocorrelation decay in velocity in the famous work of Alder and Wainright 1970.
Combining all these fittings, a close form expression for the response of the liquid is proposed which models the spatial temporal response of the fluid very well after time $t>0.5$ and when the wave has passed beyond $x>2.5$.
The error at short times is attributed to the discrete nature of the system, with jumps as the wave moves between atoms clearly shown. 
This discrete physics is not captured by the continuum wave packet as modelled by the exponential forms.
The final fitted form presents a tool to capture viscosity by running simulation for only short times and cross correlating only local bins, before using the proposed fitting form to model the long time and long distance.
In this way, expensive and long tribology, rheology or other simulations that measure viscosity of complex molcules which demand long runs or massive ensembles can be reduced to simple short runs where a small part of the temporal-spatial plot is required.
Software to obtain spatial temporal autocorrelation and fit these to the travelling wave Gaussians are provided with the manuscript to make this techniques usable for a range of problems.

\appendix
\section{Appendix}

\subsection{Assuming Independent Functions for Fitting}

Taking this fitting to standard deviation, we weight our fitting function \eq{fitted_visc} based on the coefficient obtained by fitting to the standard deviation with value $1/x^b$.
This has the effect of increasing the importance of the short distance interactions which has the lowest signal to noise ratio and de-emphasising the later times where standard deviation becomes large.

\begin{align}
\mu(x) = A \left[\alpha X_1 ( 1 - e^{-x/X_1}) + X_2 (1 - \alpha) ( 1 - e^{-x/X_2}) \right]
\label{fitted_visc}
\end{align}

 \begin{figure}[H]
\begin{subfigure}{0.48\textwidth}
 \includegraphics[width=\textwidth]{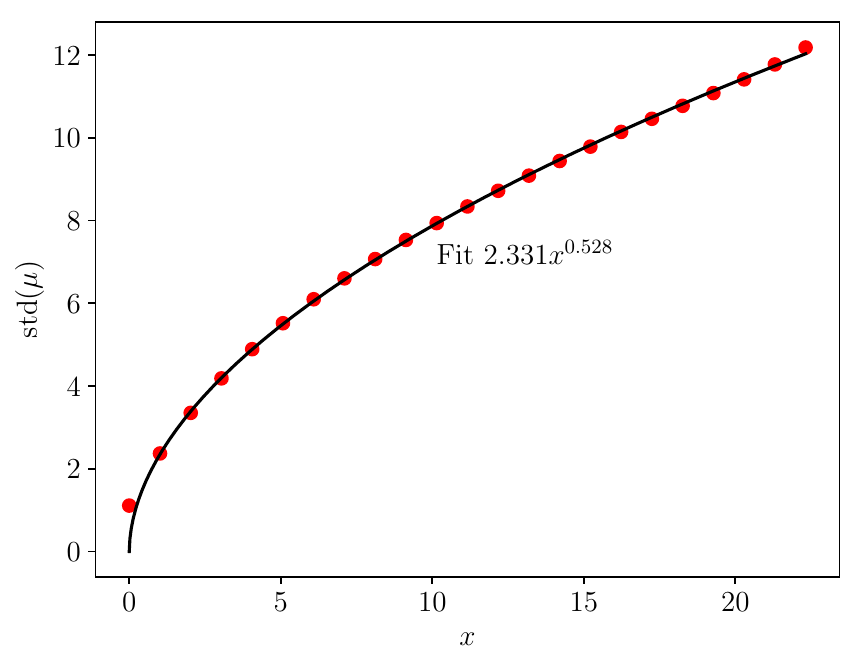}
 \caption{Standard deviation of viscosity as a function of layer with fitting of form $Ax^b$, used as weighting for the fitting function to viscosity.}  
 \label{fit_std}
\end{subfigure}
\;\;
\begin{subfigure}{0.48\textwidth}
 \includegraphics[width=\textwidth]{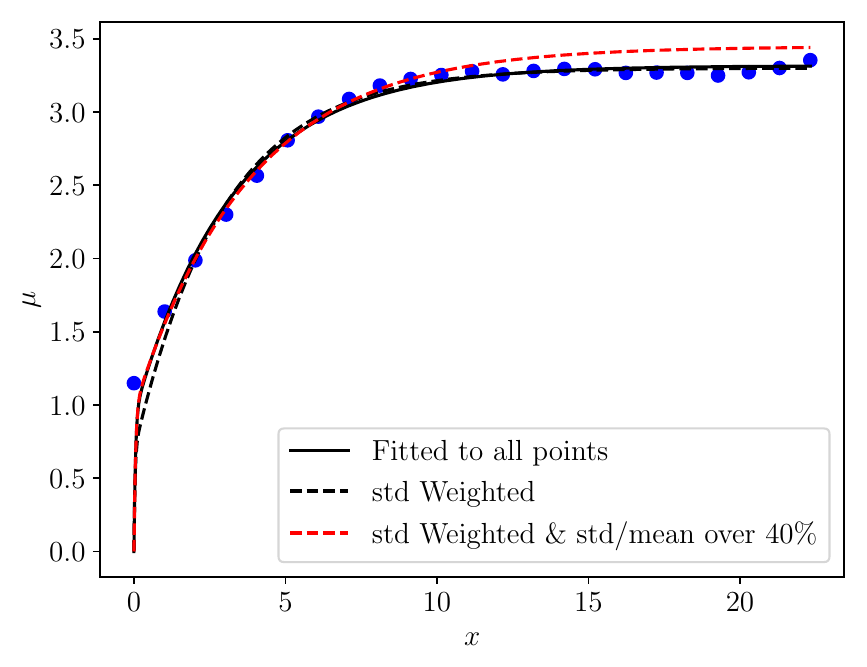}
 \caption{Resulting viscosity as a function of layer distance fitted using a combination of exponentials weighted by standard deviations}  
 \label{layer_fittings}
\end{subfigure}
\caption{Fitting process used to obtain long time viscosity measure.}
 \end{figure}

In addition, following  Zhang et al (2015) the fitted points can be truncated based on a ratio of the standard deviation to the mean value (40\% was found to be optimal for this purpose) so the long distance behaviour which is subject to noise is even further discounted.
However, the fitting function \eq{fitted_visc} is then used to obtain the infinite distance limit which is expressed in terms of the 4 fitted coefficients  $A$, $X_1$, $X_2$ and $\alpha$
\begin{align}
\mu(x \to \infty) = A \left[X_1 \alpha + X_2 (1 - \alpha)  \right]
\label{fitted_visc}
\end{align}
Combining the spatial and temporal fitting functions we can assume the space and time are independent to propose a 2D kernel,
\begin{align}
\mu(x,t) = \mu(x) \mu(t) =  A \left[\alpha X_1 ( 1 - e^{-x/X_1}) + X_2 (1 - \alpha) ( 1 - e^{-x/X_2}) \right]  \nonumber \\ 
\left[\beta \tau_1 ( 1 - e^{-t/\tau_1}) + \tau_2 (1 - \beta) ( 1 - e^{-t/\tau_2}) \right]
\label{fitted_visc}
\end{align}
Fitting this to the spatial-temporal correlation data from the SGK expressions at short time and space where standard deviation is low, we can then extrapolating to the limit to give an estimate of viscosity,
\begin{align}
\mu(x \to \infty,t \to \infty) =  A \left[\alpha X_1 + X_2 (1 - \alpha)  \right] \left[\beta \tau_1  + \tau_2 (1 - \beta)  \right]
\label{fitted_visc_limits}
\end{align}
Weighting this fit using the standard deviation variation obtained from fitting to the MD data where $std(\mu) \approx B \sqrt{x t} $.
However, this assumption of independent time and space components is clearly flawed and a much better model is provided in the main text.
The standard deviation weighted approach here is, however, extensible to this more sophisticated function.


\subsection{Other Statepoints}
\label{sec:other_statepoints}

Other statepoints including one firmly in the liquid state with $T=1$ and $\rho=0.8$ as well as one that is a solid lattice where $T=1$ and $\rho=1.05$.
In the liquid case, the fitting approach used here would clearly be applicable with similar fits expected to perform well.
The speed of sound is shown as a red line, for the liquid case this is taken from \citet{Stephen_et_al} with value of $c_s = 5.4$.
In the solid case a value of $c_s = 7$ is used as it appears to be a good fit to the data. 
For the solid, it is clear that the oscillatory decay fit set out in Figure \ref{A1_A2_fits}b) would become essential to capture the much slower decay.
The fitting process would also need to be refined to fit many Gaussians $c_1$, $c_2$, $c_3$, etc as the various molecular cages are represented in the spatial signal.
The wave also follows the speed of sound for a much longer time in the solid which suggest this departure is a consequence of the deforming liquid structure over time.
\begin{figure}[H]
\begin{subfigure}{0.45\textwidth}
    \includegraphics[width=\textwidth]{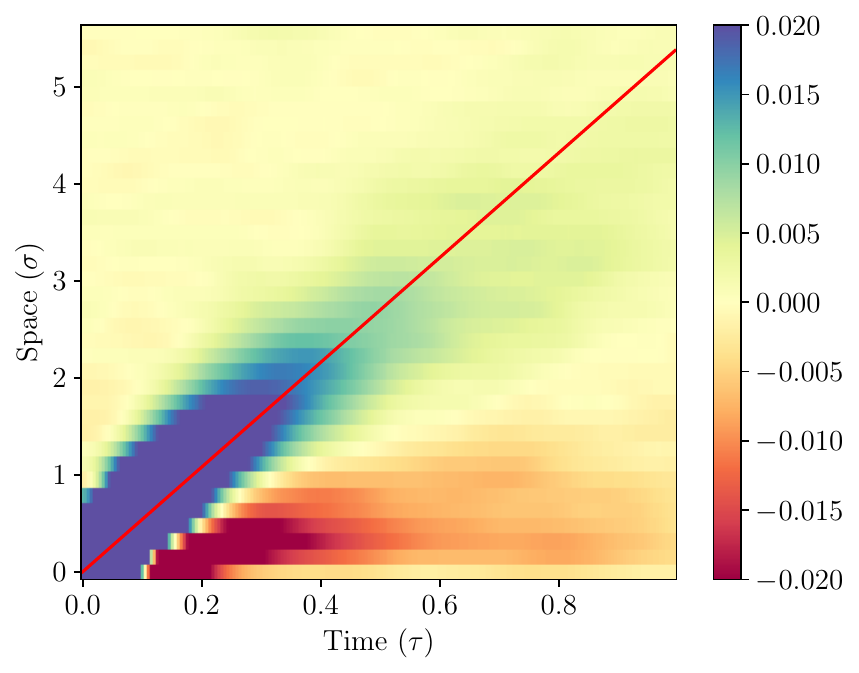}
\caption{Liquid at $T=1$ and $\rho=0.8$}
\end{subfigure}
\;\;\;\;
\begin{subfigure}{0.45\textwidth}
    \includegraphics[width=\textwidth]{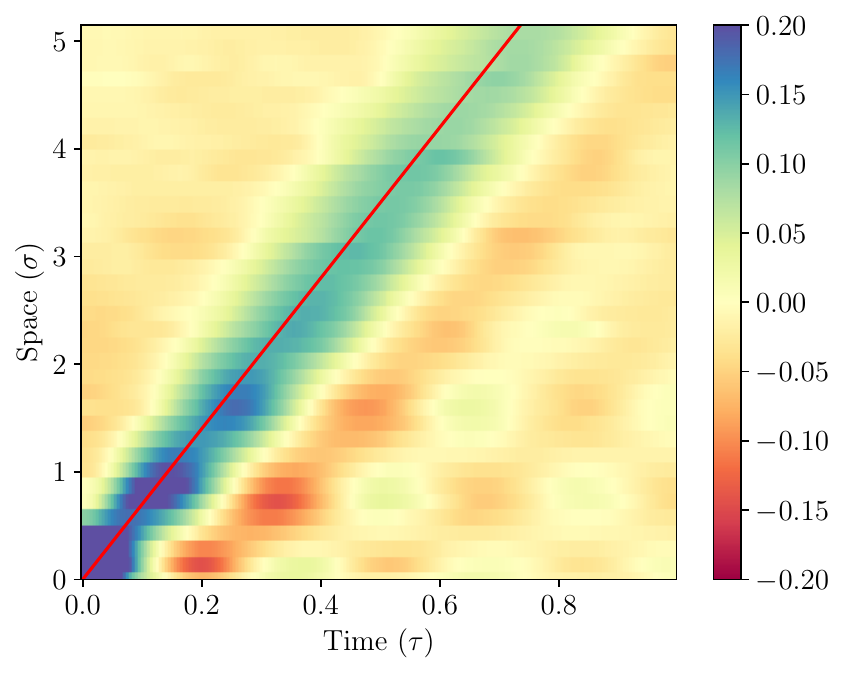}
\caption{Solid at $T=1$ and $\rho=1.05$}
\end{subfigure}
\caption{Testing the method for two different statepoints including a point clearly in the liquid phase $a)$ and a solid $b)$. NOTE these are normalised and wrong colourmap, fix them!}
\label{Other caes}
\end{figure}
These plots show the deep insight into the liquid and solid state provided by the real space extension of the Green Kubo function, even in this simple LJ system.
Application to more complex molecules would be expected to yield further insights and modelling possibilities.

\subsection{Virial Stress}

\begin{figure}
 \includegraphics[width=0.9\textwidth]{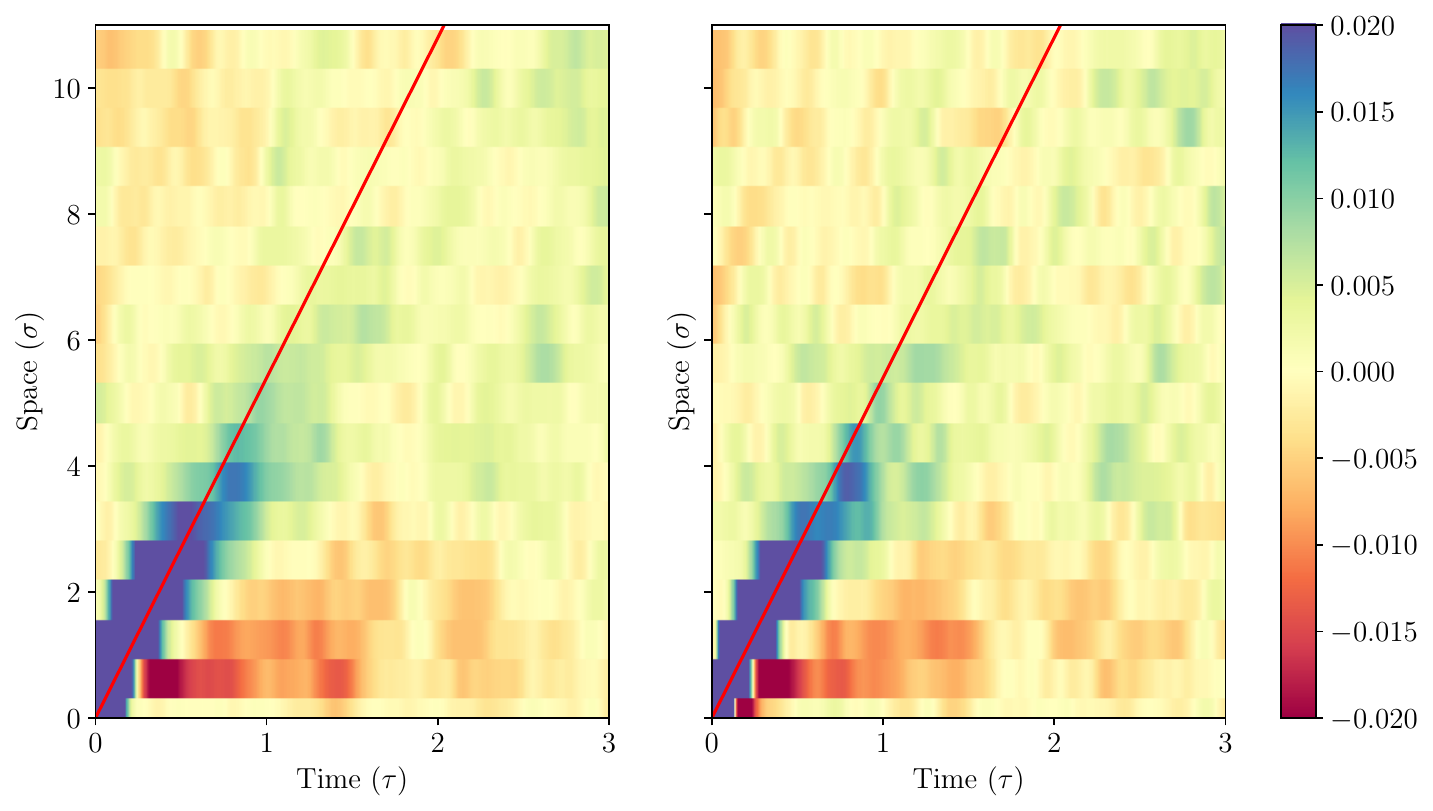}
\put(-340,190){$a)$}
\put(-180,190){$b)$}
 \caption{The resulting spatial temporal response compared for a domain decomposed using the IK1 defintion of pressure from Eq \ref{virial} in a) and the VA pressure of Eq \ref{VA} in b) (where VA was used for all other plots in this work), where 17 layers were used in both cases (35 cells total for domain). Some subtle differences are observed at short times due to the difference in assigning forces to bins, whereas difference in longer times and further in space are likely due to difference in statistical noise in the simulations}.  
 \label{IK_vs_VA}
 \end{figure}

In this section, we show the use of a virial or IK1 stress Eq \ref{virial} still produces similar spatial temporal information to the volume average of Eq \ref{VA}.
This is shown by the comparison in Figure \ref{IK_vs_VA} of the spatial and temporal response of the system.

Given the default in LAMMPS is to define ``stresslets'' to atoms and then assign to bins by chunking, this suggests the analysis presented in this work would be directly applicable to LAMMPS without having to implement a volume averages style stress based on the fraction of line inside a box.
Formally the use of decompositon of global stress into IK1 or volume average contributions in the Green Kubo relation, as in Eq. \ref{virial_to_VA}, are equally valid.
It is possible the use of the IK1 might have greater tendency to expose the atomic oscillations in the stress correlation measurements as explored in Figure \ref{short_time_oscillations} where weighting stress based on length of interactions in a volume tends rather than atomic position tends to have a smoothing effect on local stress measurements.

\vskip 10 truemm

\bibliography{./ref}

\end{document}